\documentclass[aps,prd,twocolumn,superscriptaddress,letterpaper,nofootinbib,
               nopreprintnumbers,showpacs]{revtex4-1}
\usepackage{amsmath}
\usepackage{microtype}
\usepackage{siunitx}
\usepackage{txfonts}
\usepackage{verbatim}
\usepackage{dcolumn}

\usepackage[usenames, dvipsnames, svgnames, table]{xcolor}
\usepackage[colorlinks=true, citecolor=RoyalBlue, linkcolor=BrickRed, urlcolor=ForestGreen]{hyperref}
\usepackage{graphicx}

\usepackage[justification=centerfirst]{caption}

\begin{document}
\title{Calibration Uncertainty for Advanced LIGO's First and Second Observing Runs}

\author{Craig Cahillane}
\email[]{ccahilla@caltech.edu}
\affiliation{LIGO Laboratory, California Institute of Technology, Pasadena, CA 91125, USA}
\author{Joe Betzwieser}
\affiliation{LIGO Livingston Observatory, Livingston, LA 70803, USA}
\author{Duncan A. Brown}
\affiliation{Syracuse University, Syracuse, NY 13244, USA}
\author{Evan Goetz}
\affiliation{LIGO Hanford Observatory, Richland, WA 99352, USA}
\author{Evan D. Hall}
\affiliation{LIGO Laboratory, Massachusetts Institute of Technology, Cambridge, MA 02139, USA}
\author{Kiwamu Izumi}
\affiliation{LIGO Hanford Observatory, Richland, WA 99352, USA}
\author{Shivaraj Kandhasamy}
\affiliation{LIGO Livingston Observatory, Livingston, LA 70803, USA}
\author{Sudarshan Karki}
\affiliation{University of Oregon, Eugene, OR 97403, USA}
\author{Jeff S. Kissel}
\affiliation{LIGO Hanford Observatory, Richland, WA 99352, USA}
\author{Greg Mendell}
\affiliation{LIGO Hanford Observatory, Richland, WA 99352, USA}
\author{Richard L. Savage}
\affiliation{LIGO Hanford Observatory, Richland, WA 99352, USA}
\author{Darkhan Tuyenbayev}
\affiliation{University of Texas Rio Grande Valley, Brownsville, TX 78520, USA}
\author{Alex Urban}
\affiliation{LIGO Laboratory, California Institute of Technology, Pasadena, CA 91125, USA}
\author{Aaron Viets}
\affiliation{University of Wisconsin-Milwaukee, Milwaukee, WI 53201, USA}
\author{Madeline Wade}
\affiliation{Kenyon College, Gambier, OH 43022, USA}
\author{Alan J. Weinstein}
\affiliation{LIGO Laboratory, California Institute of Technology, Pasadena, CA 91125, USA}

\begin{abstract}
Calibration of the Advanced LIGO detectors is the quantification of the detectors' response to gravitational waves. 
Gravitational waves incident on the detectors cause phase shifts in the interferometer laser light which are read out as intensity fluctuations at the detector output. 
Understanding this detector response to gravitational waves is crucial to producing accurate and precise gravitational wave strain data.
Estimates of binary black hole and neutron star parameters and tests of general relativity require well-calibrated data, as miscalibrations will lead to biased results.
We describe the method of producing calibration uncertainty estimates for both LIGO detectors in the first and second observing runs.
\end{abstract}

\pacs{04.30.-w, 04.80.Nn, 95.55.Ym}

\maketitle

\section{Introduction}
The Laser Interferometer Gravitational-wave Observatory (LIGO), with its twin detectors in Hanford, Washington (H1) and Livingston, Louisiana (L1) has directly observed transient gravitational wave (GW) signals \cite{GW150914, GW151226, GW170104}.
These events are consistent with binary black hole coalescences \cite{O1BBHPaper}, whose detections have ushered in a new era of gravitational wave astronomy.
Observing Run 1 (O1) saw the first Advanced LIGO GW strain data taken between September 18th, 2015 through January 12th, 2016.  
Observing Run 2 (O2) started on November 30th, 2016, and ended August 25th, 2017.  

GW signals are extremely rich sources of information from previously unexplored astrophysical phenomena.  
The uncertainty in the estimated amplitude and phase of the GW directly impacts the astrophysics we can learn from both transient and long-duration signals.
For compact binary coalescence GW signals, estimates of the progenitor masses, spins, luminosity distance, orbital plane inclination, final mass, and sky location are derived from the detected waveforms, and each are potentially limited by calibration accuracy \cite{GW150914PEPaper}.
The rate at which such systems coalesce in the universe can be drawn from detected events, but as the number of observations increases, rate estimates will become limited by strain amplitude uncertainty \cite{RatesPaper, GW170104}. 
Testing general relativity has begun with the first detections \cite{GW150915GRTests, GW170104}, but as the detectors' sensitivity improves and there are more high signal-to-noise ratio events, calibration uncertainty will limit our test results, and calibration error will bias our test results \cite{TestingGR2016, TestingGR2017}.
Upper limits and observations of sources of continuous gravitational waves, such as rapidly rotating neutron stars, depend on calibration uncertainty \cite{O1CW2017a, O1CW2017b}.
Upper limits and observations of the GW stochastic background of unresolvable sources depend on the amplitude calibration uncertainty \cite{O1Stoch2017a, O1Stoch2017b}.
Using many GW detections to refine estimates of the Hubble constant will be fundamentally limited by calibration uncertainty \cite{Chernoff1993, Cosmology}. 

The total calibration uncertainty budget consists of statistical uncertainty and systematic error.
Statistical uncertainty is the intrinsic uncertainty associated with measurements.
Systematic error is the bias quantifying the difference between model and measurement.
These quantities will be further defined in Section \ref{sec:Detectors}.

This paper presents a refined method of producing calibrated GW strain data error and uncertainty budgets, discusses the error and uncertainty budget's evolution over time throughout the first two observing runs, and highlights the results relevant to the subsequent transient detections.
The work here builds on previous work presented for the detection of GW150914 \cite{GW150914CalPaper}. 
Section \ref{sec:Detectors} reviews the detector fundamentals, how the strain time-series $h(t)$ is constructed from the detector output, the model of the detector response used to construct that estimate, and measurements supporting that model.
Section \ref{sec:Method} explains how the statistical uncertainty and systematic error budget on the strain time-series is constructed.
Section \ref{sec:Results} shows error and uncertainty budgets for observational data sets to date, with focus on the times of the GW detections.
Section \ref{sec:FutureWork} presents ideas for future work to further improve the calibration uncertainty.
Section \ref{sec:Conclusion} discusses the implications of the calibration uncertainty results.

\section{Detectors} 
\label{sec:Detectors}

The Advanced LIGO detectors are Michelson interferometers whose arms are enhanced with 4 km long Fabry-P\'erot resonant cavities.
The cavities are filled with continuous carrier laser light from an Nd:YAG 1064nm laser. 
Additional recycling cavities at the Michelson's input and output ports further improve the detector sensitivity to GWs \cite{AdvLIGOPaper, GW150914DetectorPaper}. 

A gravitational wave incident on the detector modifies the distance between the input and end mirrors of the arm cavities. 
This causes an apparent differential change in length of two arms, $\Delta L_{\text{free}}$, relative to the average length of the arms, $L$. 
Differential arm (DARM) displacement is defined as 
\begin{align}
\label{DeltaDARM} \Delta L_{\text{free}} = \Delta L_\text{x} - \Delta L_\text{y} \equiv h L,
\end{align}
where $\Delta L_\text{x}$ and $\Delta L_\text{y}$ are changes in the X and Y arm lengths, and $h$ is the detector's reconstructed GW strain signal and contains both the desired astrophysical information and unwanted noise.
We report precision calibration for gravitational waves in the frequency band between 10 and 5000 Hz.

\begin{figure}
\begin{center}
\includegraphics[width=1.0\columnwidth] {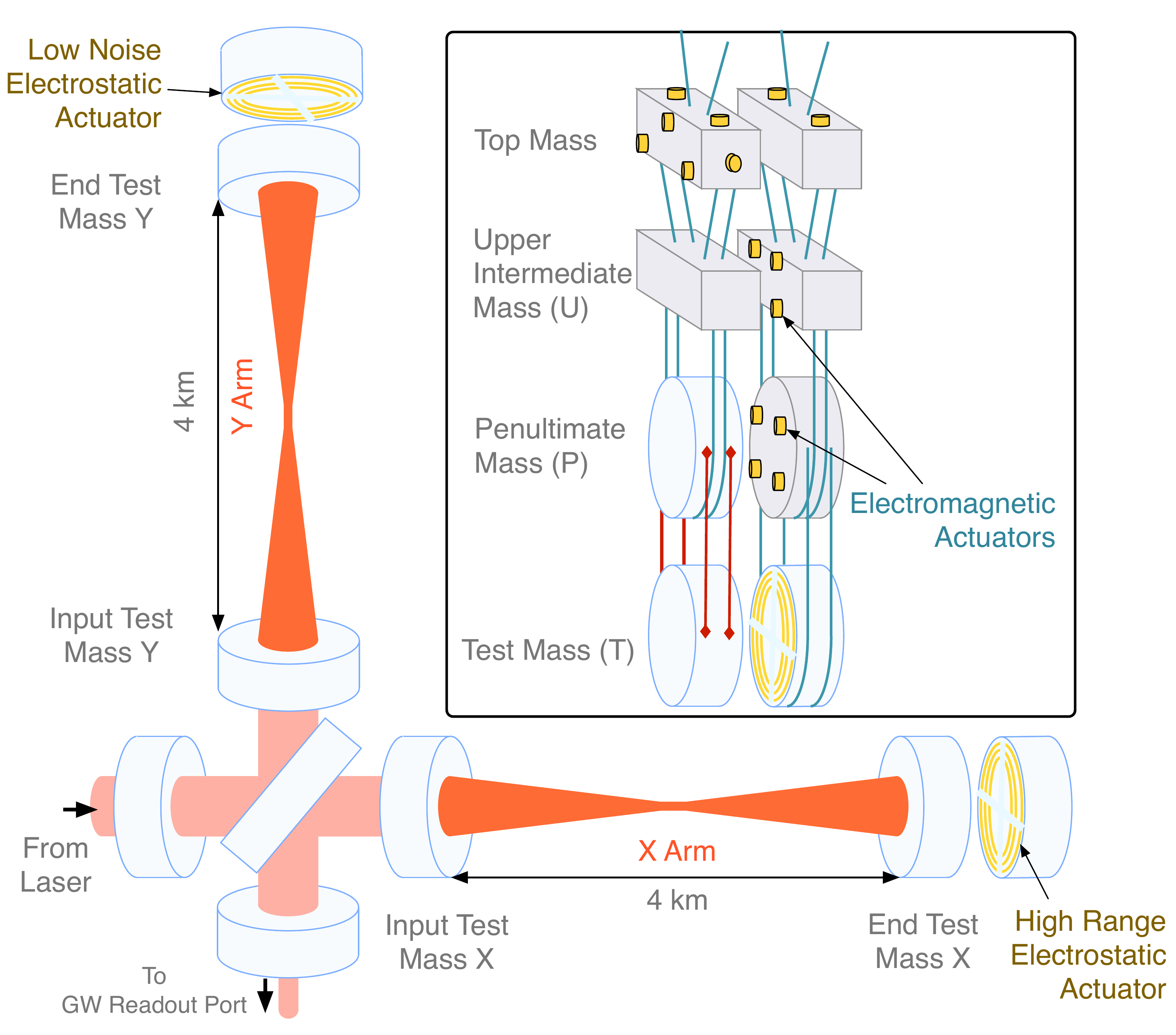}
\caption{Simplified Advanced LIGO Interferometer Layout.\\
Laser light enters the interferometer at the lower left through a power-recycling mirror, and is split by a 50/50 beamsplitter.
Two 4 km long Fabry-P\'erot resonating arm cavities are formed from four highly reflective test masses.
Laser light builds up in the cavities, reaching about 100 kW of laser power in O1 and O2.
A signal-recycling mirror between the beamsplitter and GW readout photodetector modifies the detector response to increase the detector bandwidth.
Inset: one of the quadruple pendulum suspension systems which holds each of the four test masses is shown.
}
\label{fig:IFODiagram}
\end{center}
\end{figure}

DARM displacement generates laser power fluctuations on the antisymmetric port photodetector, shown by the GW readout port in Figure \ref{fig:IFODiagram} \cite{aLIGOLSCPaper}.
When the interferometer DARM degree of freedom is held on resonance, light in the antisymmetric port destructively interferes, and laser power fluctuations on the antisymmetric photodetector are quadratic in strain. 
``On resonance'' means the round-trip DARM length is an integer number of laser wavelengths. 
A small length offset is introduced such that the interference of the laser beams from two cavities is not completely destructive in the direction of the antisymmetric photodetector. 
With the offset, laser power fluctuations are approximately proportional to strain, allowing us to directly read out incident GW strain from the laser fluctuations. 

The arm cavity mirrors are suspended from multi-stage cascading pendula \cite{Suspensions2002,Suspensions2012} and active seismic isolation systems \cite{aLIGOSEI} to suppress DARM displacement from ground motion and other force noise.
Still, DARM displacement must be further controlled to hold the interferometer on resonance.
This requires a global arm length control system, with open loop transfer function $G$, to suppress the free displacement $\Delta L_{\text{free}}$ to a smaller residual displacement via actuators present on the cascaded pendula \cite{QUADSensorsAndActuators} to a residual differential arm length $\Delta L_{\text{res}}$,
\begin{align}
\Delta L_{\text{res}} = \Delta L_{\text{free}} / [1 + G].
\end{align}

\begin{figure}[h!]
\begin{center}
\includegraphics[width=1.0\columnwidth] {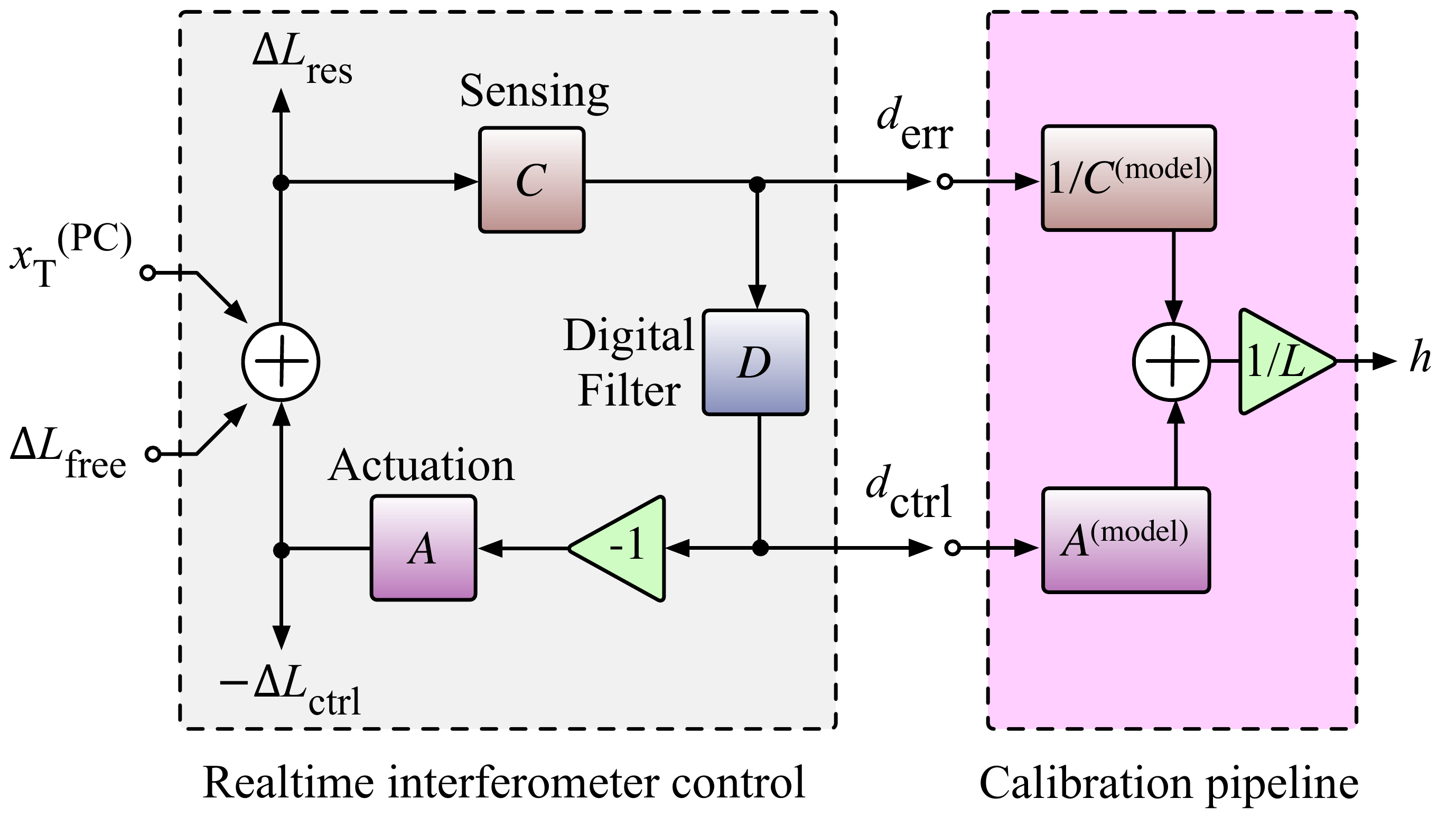}
\end{center}
\caption{DARM Control Loop and Calibration Procedure.\\
The DARM control loop is shown in the grey box on the left.
The sensing plant $C$ produces the detector output $d_{\text{err}}$ in linear response to residual differential arm motion $\Delta L_{\text{res}}$.
The digital filters $D$ are known filters conditioning the detector output $d_{\text{err}}$ into a control signal $d_{\text{ctrl}}$.
The actuation plant $A$ takes the control signal $d_{\text{ctrl}}$ and actuates on the optics by $\Delta L_{\text{ctrl}}$ to maintain cavity resonance.
The pink box on the right shows the calibration procedure, consisting of an inverse sensing model $1/C^{\text{(model)}}$ and actuation model $A^{\text{(model)}}$.
The output of the calibration pipeline is GW strain data $h(t)$.
}
\label{fig:DARMControlLoop}
\end{figure}

We define three independently quantifiable transfer functions of the DARM control loop, shown schematically in Figure \ref{fig:DARMControlLoop}. 
The sensing function $C = d_{\text{err}} / \Delta L_{\text{res}}$ defines the measured interferometric laser power response to DARM displacement and the digitization process of the power fluctuations to form the digital error signal $d_{\text{err}}$. 
Digital filters $D = d_{\text{ctrl}} / d_{\text{err}} $ convert the loop error signal to the loop control signal.
The actuation function $A = \Delta L_{\text{ctrl}} / d_{\text{ctrl}}$ generates force on the optical cavity pendula to largely cancel any detected DARM displacement within the DARM loop bandwidth. 
All transfer functions are complex-valued functions of frequency, with quantifiable magnitude and phase.
The digital filters $D$ shape the DARM loop frequency response and are known to negligible uncertainty.  
The DARM loop transfer functions $C$ and $A$ must be measured and modeled in the frequency domain between 5 and 5000 Hz.
Both $C$ and $A$ contribute to the total calibration uncertainty budget. 

The digital signals $d_{\text{err}}$ and $d_{\text{ctrl}}$ are digitally filtered to form a time-series estimate of the GW strain $h(t)$ used for astrophysical searches. 
The digital filters applied to $d_{\text{err}}$ and $d_{\text{ctrl}}$ are constructed from models of the sensing function $C^{(\text{model})}$ and actuation function $A^{(\text{model})}$:
\begin{align}
\label{eq:hoft} h = \frac{1}{L} \left[\frac{1}{C^{(\text{model})}} * d_{\text{err}} + A^{(\text{model})} * d_{\text{ctrl}}\right],
\end{align}
where $*$ indicates convolution in the time domain, or multiplication in the frequency domain, and $L$ is the  length of the interferometer arms, known to negligible uncertainty.
The accuracy and precision of the models $C^{(\text{model})}$ and $A^{(\text{model})}$ define the systematic error and statistical uncertainty in the estimated time series $h(t)$. 

We define a transfer function called the response function $R$,
\begin{align}
\label{eq:ResponseDef} h = R * d_{\text{err}} = \frac{1}{L} \left( \frac{1 + G}{C} \right) d_{\text{err}}  
\end{align}
where the DARM open loop gain $G = C * D * A$.
Equation \ref{eq:ResponseDef} illustrates that in the frequency domain, response function error $\delta R$ is equivalent to the GW strain data error $\delta h$ and response function uncertainty $\sigma_{R}$ is equivalent to the GW strain data uncertainty $\sigma_{h}$.
Throughout this paper, the response error and uncertainty relative to the calibration pipeline model $R^{(\text{model})}$ are quantified as a function of frequency $f$ with time dependence $t$:
\begin{align}
\label{eq:Response2DataEquiv} \dfrac{\delta R(f, t)}{R^{(\text{model})}} = \dfrac{\delta h (f, t)}{h}, \,\,\,\,\,\,\,\, \dfrac{\sigma_{R}(f, t)}{R^{(\text{model})}} = \dfrac{\sigma_{h}(f, t)}{h}.
\end{align}

The DARM loop transfer functions $C$ and $A$ are measured and modeled in the frequency domain. 
Additionally, the values of $C$ and $A$ can drift slowly over time, giving functions of frequency that vary in time $C(f, t)$ and $A(f, t)$.
However, our online calibration pipeline digital filters $1/C^{(\text{model})}$ and $A^{(\text{model})}$ are not perfect representations of our understanding of the interferometer.  
This leads to known systematic errors in our $h(t)$ reconstruction, governed by the sensing and actuation systematic errors $\delta C(f, t)$ and $\delta A(f, t)$.
The systematic errors relative to $C^{(\text{model})}$ and $A^{(\text{model})}$ are quantified as
\begin{align}
\label{eq:TFSystErrors}\dfrac{\delta C(f, t)}{C^{(\text{model})}} = \dfrac{C(f, t)}{C^{(\text{model})}} -1, \,\,\,\,\,\,\,\, \dfrac{\delta A(f, t)}{A^{(\text{model})}} = \dfrac{A(f, t)}{A^{(\text{model})}}-1,
\end{align}
where $C(f, t)$ and $A(f, t)$ represent the measured sensing and actuation transfer functions.

Systematic errors $\delta C$ and $ \delta A$ propagate forward to the relative response function systematic error $\delta R / R^{(\text{model})}$:
\begin{align}
\label{eq:RespSystErrors} \nonumber \dfrac{\delta R(f, t)}{R^{(\text{model})}} &= \dfrac{R(f,t)}{R^{(\text{model})}} - 1 = \left( \dfrac{1 + G(f,t)}{C(f, t)}\right) \Bigg/ \left(\dfrac{1 + G^{(\text{model})}}{C^{(\text{model})}}\right) - 1 \\
&= \dfrac{ \left( G^{(\text{model})} \, \dfrac{\delta A(f, t)}{A^{(\text{model})}} - \dfrac{\delta C(f, t) / C^{(\text{model})} }{1 + \delta C(f, t) / C^{(\text{model})}} \right) }{1+G^{(\text{model})}}.
\end{align}

The response function uncertainty $\sigma_R(f,t)$ is in general a $2 \times 2$ matrix to represent uncertainty in the complex plane.
The off-diagonal terms are capable of capturing covariance between the two basis vectors.
In this paper, we will be using Bendat and Piersol relative magnitude uncertainty and absolute phase uncertainty, seen in Equation \ref{eq:CoherenceUncertainty} \cite{BendatPiersolCoherenceUncertainty}.
All uncertainties will be propagated in the relative magnitude and absolute phase basis with no covariance, as given from Bendat and Piersol.

In previous calibration uncertainty work \cite{GW150914CalPaper}, covariance between actuation stages was found to be non-negligible and was included in the uncertainty budget.
In this work, improved measurement techniques broke this covariance, rendering its effect negligible.
There is also no covariance between the sensing function and any actuation stage.

The $2 \times 2$ uncertainty matrices $\sigma_{C}(f,t)$ and $\sigma_{A}(f,t)$ propagate to the relative response function uncertainty $\sigma_{R} / R^{(\text{model})}$:
\begin{align}
\label{eq:RespUnc} \nonumber \dfrac{\sigma_{R}(f, t)}{R^{(\text{model})}} &= \dfrac{1}{R^{(\text{model})}} \sqrt{ \left( \dfrac{\partial R}{\partial C}\right)^2 \sigma_C^2 + \left( \dfrac{\partial R}{\partial A}\right)^2 \sigma_A^2 } \\
&= \dfrac{C^{(\text{model})}}{1 + G^{(\text{model})}}\sqrt{ \dfrac{1}{C(f,t)^4} \, \sigma_C(f,t)^2 + D(f)^2 \, \sigma_A(f,t)^2}.
\end{align}

Together $\delta R / R^{(\text{model})}$ and $\sigma_{R} / R^{(\text{model})}$ make up the entire calibration error and uncertainty budget.

\subsection{Sensing Model}
\label{subsec:SensingModel}
\begin{figure*}
\includegraphics[width=0.48\textwidth]{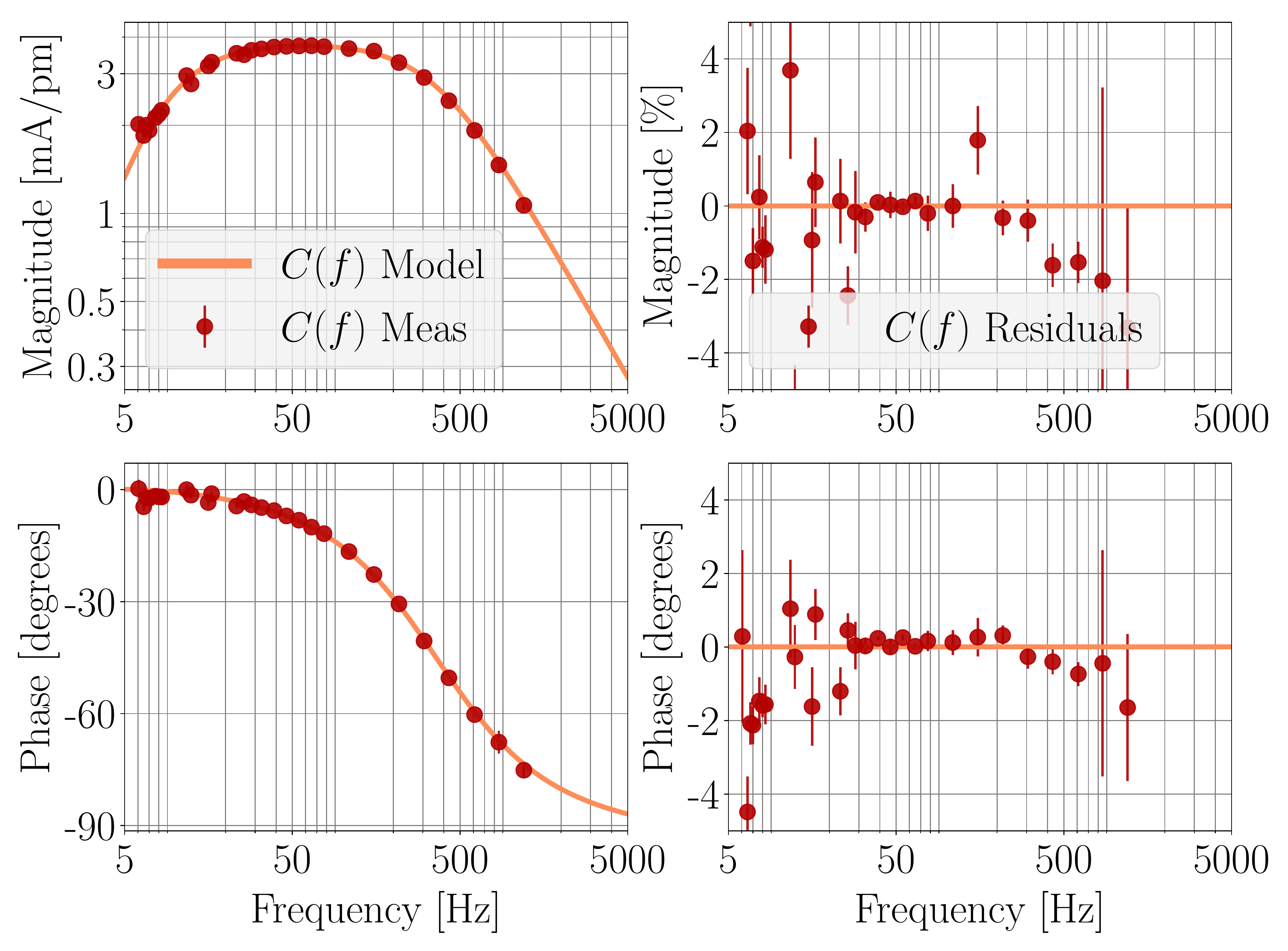}
\includegraphics[width=0.48\textwidth]{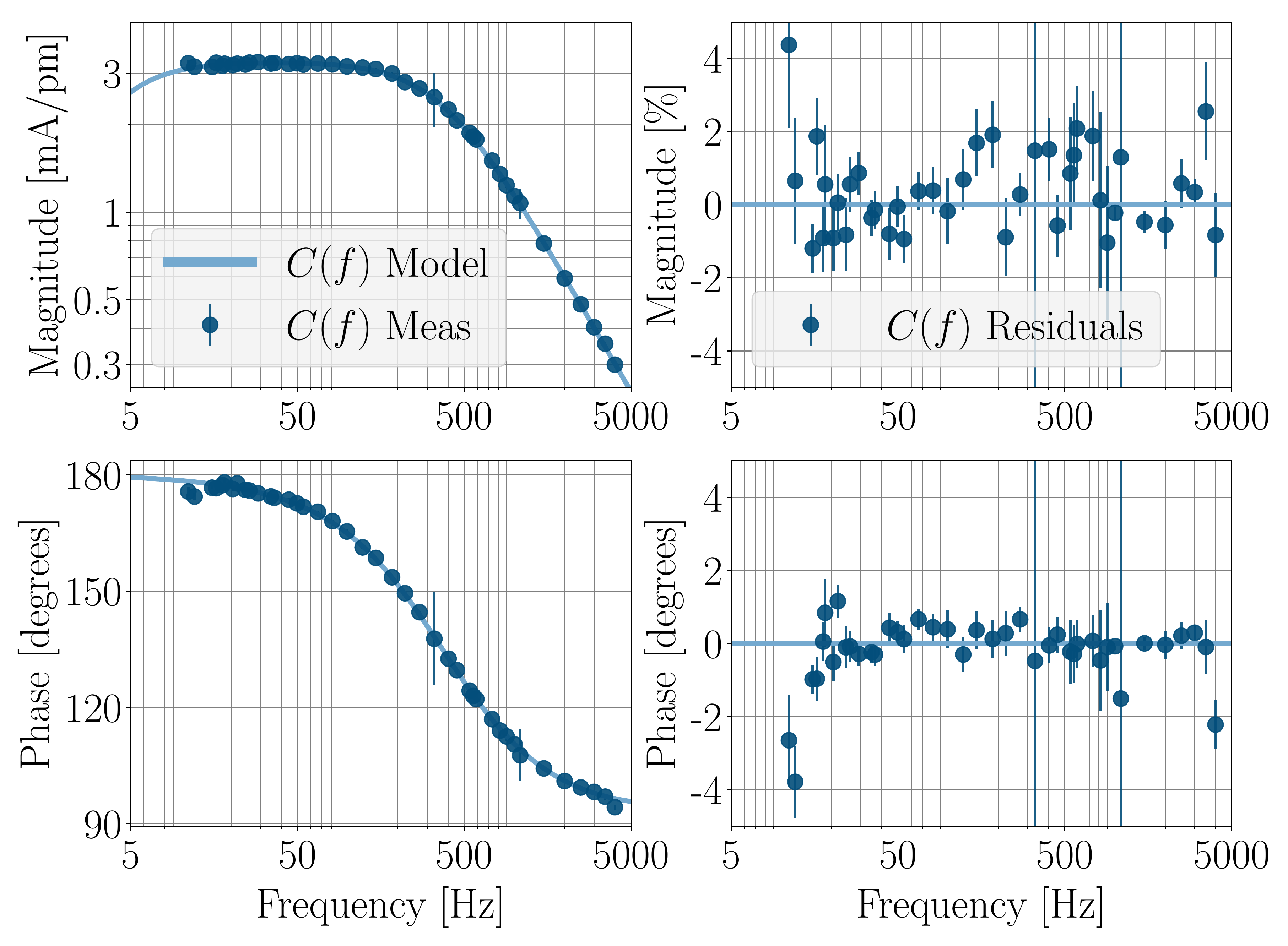}
\caption{Sensing measurements $C^{\text{(meas)}}(f)$, sensing model $C^{(\text{model})}(f,t,\vec{\lambda}_{C})$, and their residuals $\delta C(f, t)/C^{(\text{model})}(f,t,\vec{\lambda}_{C})$.  \\
The H1 sensing reference measurement from January 4th, 2017 is shown in the four panels on the left in red. 
The L1 sensing reference measurement from November 26th, 2016 is in the four right panels in blue.
The first and third columns are the Bode plots of $C$, while the second and fourth columns are the residuals.
The model parameters $\vec{\lambda}_{C}$ were found via an MCMC.
Physically, the magnitude Bode plots represent how many milliamps of current are generated at our transimpedence photodetector per picometer of differential arm motion from 5 to 5000 Hz.
The drop in sensitivity at low frequencies shows the effect of detuning at both detectors.
The 180 degree phase difference between H1 and L1 is a sign convention difference between the detectors.
}
\label{fig:SensingPlant}
\end{figure*}

The core of sensing function model $C^{\text{(model)}}$ is the interferometric transfer function from DARM displacement to laser power on the antisymmetric port photodetectors.
The photodetectors generate photocurrent in response, which is then run through a transimpedence amplifier, whitening filters, anti-aliasing filters, and an analog-to-digital converter to produce DARM error counts $d_{\text{err}}$.
The transfer function from DARM displacement in picometers to milliamps of photodetector output current is shown in Figure \ref{fig:SensingPlant}.

Advanced LIGO's dual-recycled, Fabry Perot Michelson detectors operate in the ``resonant signal extraction'' configuration:
the signal recycling mirror is purposefully detuned from resonance to increase the bandwidth of the detector.
When detuned exactly to 90 degrees, as designed, the sensing model may be approximated by as a single pole system.
This model was used for estimating the uncertainty and error of GW150914 \cite{GW150914CalPaper}.
However, measurements have revealed both detectors' signal recycling cavities are slightly offset from 90 degrees detuning, inducing an optical anti-spring which reduces the displacement response at low frequencies. 

We employ another approximation to the interferometric response in the sensing model, $C^{(\text{model})}(f,t,\vec{\lambda}_{C})$ \cite{BuonnanoChen2002, WardThesis, HallThesis}.  
The model contains all terms from \cite{GW150914CalPaper} but additionally includes an optical anti-spring term defined by $f_{S}$ and $Q_{S}$, the optical anti-spring pole frequency and quality factor,
\begin{eqnarray}
\label{eq:SensingModel} C^{(\text{model})}(f,t,\vec{\lambda}_{C}) & = & \dfrac{\kappa_C(t) \, H_C}{1+i f / f_{CC}} \, C_{R}(f)  \, e^{-2 \pi i f \tau_{C}} \nonumber\\
& & \times \,\, \dfrac{f^2}{f^2 + f_S^{2} - i f f_S Q_S^{-1}} 
\end{eqnarray}
The optical gain $H_C$ defines the scale of the sensing function in units of error signal counts $d_{\text{err}}$ per DARM displacement in meters. 
It collects all individual scale factors from the interferometric response in watts / meter, the optical efficiency of the photodiodes in amps / watt, through the transimpedance analog electronics in volts / amp, and recorded in analog-to-digital converter counts / volt.
The time dependent scale factor $\kappa_{C}(t)$, initially set to 1 at the reference time, accounts for slow changes in $H_C$ as the detector's interferometric response evolves due to mirror alignment drift and thermal loading \cite{CALTimeDependence}.
The coupled cavity pole $f_{CC}$ defines the detector bandwidth.
The sensing time delay $\tau_{C}$ includes the light travel time over the length of the arms, computational delay in the digital acquistion system, and a compensation for the exclusion of additional, high-frequency response of the Fabry-P\'erot arm cavities beyond the single coupled cavity pole model \cite{Rakhmanov2002}.
The model time delay $\tau_{C}$ is 77.6 $\mu$s for both detectors.
The frequency dependent function $C_{R}(f)$ is the response of the digital acquisition system, including transimpedance electronics and anti-aliasing filters, all known to negligible uncertainty.
The parameter vector $\vec{\lambda}_C$ defines a set of the time-independent, reference sensing parameters whose values are fit to non-negligible precision: 
$\vec{\lambda}_{C} = \begin{pmatrix} H_C & f_{CC} & \delta \tau_{C} & f_{S} & Q_{S}^{-1} \end{pmatrix}^T$, where $\delta \tau_{C}$ is a correction time delay factor on the model time delay $\tau_{C}$.
The nominal values of the reference sensing parameters $\vec{\lambda}_{C}$ for each detector are found in Table \ref{tab:SensingTable}.

Our model of the sensing function $C^{(\text{model})}(f,t,\vec{\lambda}_{C})$ is an approximation.
The true detector sensing function changes over time and deviates from the sensing model at high frequencies.
The sensing model dynamically corrects for $\kappa_{C}(t)$ with real-time measurement. 
However, $f_{CC}$, $f_{S}$, and $Q_{S}^{-1}$ are also changing in time, but are not corrected for in the model.
At present, the time dependence in $f_{CC}$ is included in the calibration uncertainty budget as a known systematic error, since it is tracked via real-time measurement but cannot yet be dynamically corrected for in the model.
The time dependence in $f_{S}$ and $Q_{S}^{-1}$ results in expanded uncertainty at low frequency.
The total systematic error in the sensing function, $\delta C(f, t)$, is
\begin{align}
\label{eq:SensingError} \dfrac{\delta C(f, t)}{C^{(\text{model})}} = \left( \frac{1 + i f / f_{CC}}{1 + i f / f_{CC}(t)} \right) \, \dfrac{\delta C^{GP}(f)}{C^{(\text{model})}} \, e^{-2 \pi i f \delta \tau_{C}}.
\end{align}
The first term is the explicit correction for time dependence of the coupled cavity pole, $f_{CC}(t)$.
A correction time delay factor $\delta \tau_{C}$ modifies the original time delay $\tau_{C}$ included in the model.
Further systematic errors may originate from the uncorrected time dependence of $f_{S}$ and $Q_{S}^{-1}$ or additional unknown systematic errors. 
Any remaining frequency dependent systematic errors are covered by a Gaussian Process regression $\delta C^{GP}(f)$.  
Quantifying errors $\delta C^{GP}(f)$ is explained further in Section \ref{sec:Method}.

\subsection{Actuation Model}
\label{subsec:ActuationModel}

The Advanced LIGO test masses are suspended via quadruple cascaded pendula. 
Each suspension stage has independent actuators, as shown in Figure \ref{fig:IFODiagram}. 
The control signal, $d_{\text{ctrl}}$, is digitally distributed as a function of frequency to each stage's actuators via a digital-to-analog converter and signal processing electronics to create the control displacement, $\Delta L_{\text{ctrl}}$.
The distribution filters are designed taking into account all actuators' authority to displace the test mass.
On the upper intermediate and penultimate stage, the digital-to-analog converter drives electromagnets on the reaction stage creating a force on magnets attached to the suspended stage.
On the test mass stage, the digital-to-analog converter drives an electrostatic system which creates a force, quadratic in the applied potential, via dipole-dipole interactions between the test mass and a pattern of electrodes on the reaction mass (see Figure \ref{fig:IFODiagram}).
With a large bias voltage and low control voltage, the requested actuation forces on the electrostatic system are in the linear regime.
Any time-dependent change to the slope of the linear response due to quadratic terms is measured continuously, as described below.

The sum of the paths the digital control signal, $d_{\text{ctrl}}$, takes through each stage to displace the test mass, $\Delta L_{\text{ctrl}}$, makes up our total actuation model:
\begin{eqnarray}
\label{eq:ActuationModel} A^{(\text{model})}(f,t, \vec{\lambda}_{A}) & = & \bigg[\kappa_{T}(t) \, F_{T}(f) \, H_{T} \, A_{T}(f) \nonumber\\
& & \hspace{0.0cm} ~+~ \kappa_{PU}(t) \, \Big( F_{P}(f) \, H_{P}  \, A_{P}(f) \nonumber\\
& & \hspace{0.2cm} ~+~ F_{U}(f) \, H_{U} \, A_{U}(f) \Big) \bigg] \, e^{-2\pi i f \tau_{A}}
\end{eqnarray}
where $U$, $P$, and $T$ represent the three stages used for control; the upper-intermediate, penultimate, and test mass stages, respectively.  
Each stage is composed of the normalized electro-mechanical frequency response of the pendulum and its actuators, $A_{i}(f)$, the digital distribution filter, $F_{i}(f)$, a dimensionful scale factor, $H_{i}$, and an overall digital delay, $\tau_{A}$, defined by the common computational delay from each stage. 
The model time delay $\tau_{A}$ is 45 $\mu$s for L1 and 61 $\mu$s for H1.
$\kappa_{PU}(t)$ is the time dependence of the penultimate and upper intermediate scale factor, and $\kappa_{T}(t)$ is the time dependence of the test mass scale factor \cite{CALTimeDependence}.

The penultimate and upper intermediate scale factor $\kappa_{PU}(t)$ is not expected to vary much over time, as it represents the change in the electromagnetic coil actuators' strength.
The test mass scale factor $\kappa_{T}(t)$ does vary significantly over time as the electric charge on the test mass builds up, changing the actuation strength of the electrostatic drive.

The reference scale factor for each stage, $H_{i}$, collects scale factors from that of the digital-to-analog converter in volts / count, each stage's drive electronics in amps / volt or volts / volt, the actuator itself in newtons / amp or newtons / volt depending on the stage, and the stiffness of the suspension in meters / newton. 
Time delay correction factors for each stage $\delta\tau_{i}$ are extracted from measurements as stage-specific corrections to the overall actuation delay $\tau_{A}$.
The electro-mechanical transfer functions, $A_{i}$, for each stage are independently measured and included in the model with negligible uncertainty.
Remaining scale factor and delay parameters dominate the actuation function uncertainty, and are thus collected in the set of reference parameters $\vec{\lambda}_{A} = \begin{pmatrix}H_{U} & \delta\tau_{U} & H_{P} & \delta\tau_{P} & H_{T} & \delta\tau_{T} \end{pmatrix}^T$.
The values of these reference parameters $\lambda_{A}$ are found in Table \ref{tab:ActuationTable}.

The digital filters, $F_{i}$, are known a priori, and time-dependent corrections $\kappa_{PU}$ and $\kappa_{T}$ are dynamically corrected for when estimating $h(t)$.
The remaining components of the actuation stage model, $[H_{i}\, A_{i}]^{(\text{model})}(f, \vec{\lambda}_{i})$, may contain systematic errors.
We allow for and quantify systematic errors in each actuation stage as
\begin{align}
\label{eq:ActuationError} \dfrac{\delta A_{i}(f)}{A_{i}^{(\text{model})}} = \dfrac{\delta A_{i}^{GP}(f)}{A_{i}^{(\text{model})}} \, e^{-2 \pi i f \delta\tau_{i}}
\end{align}
where $\delta\tau_i$ is a time delay phase error on each stage, and $\delta A_{i}^{GP}(f)$ is the systematic error in scale or frequency dependence from the Gaussian process regression done on each stage's measurement residuals.
Systematic error calculations are explained fully in Section \ref{sec:Method}, subsection \ref{subsec:GP}.

\begin{figure*}
\includegraphics[width=1.0\columnwidth] {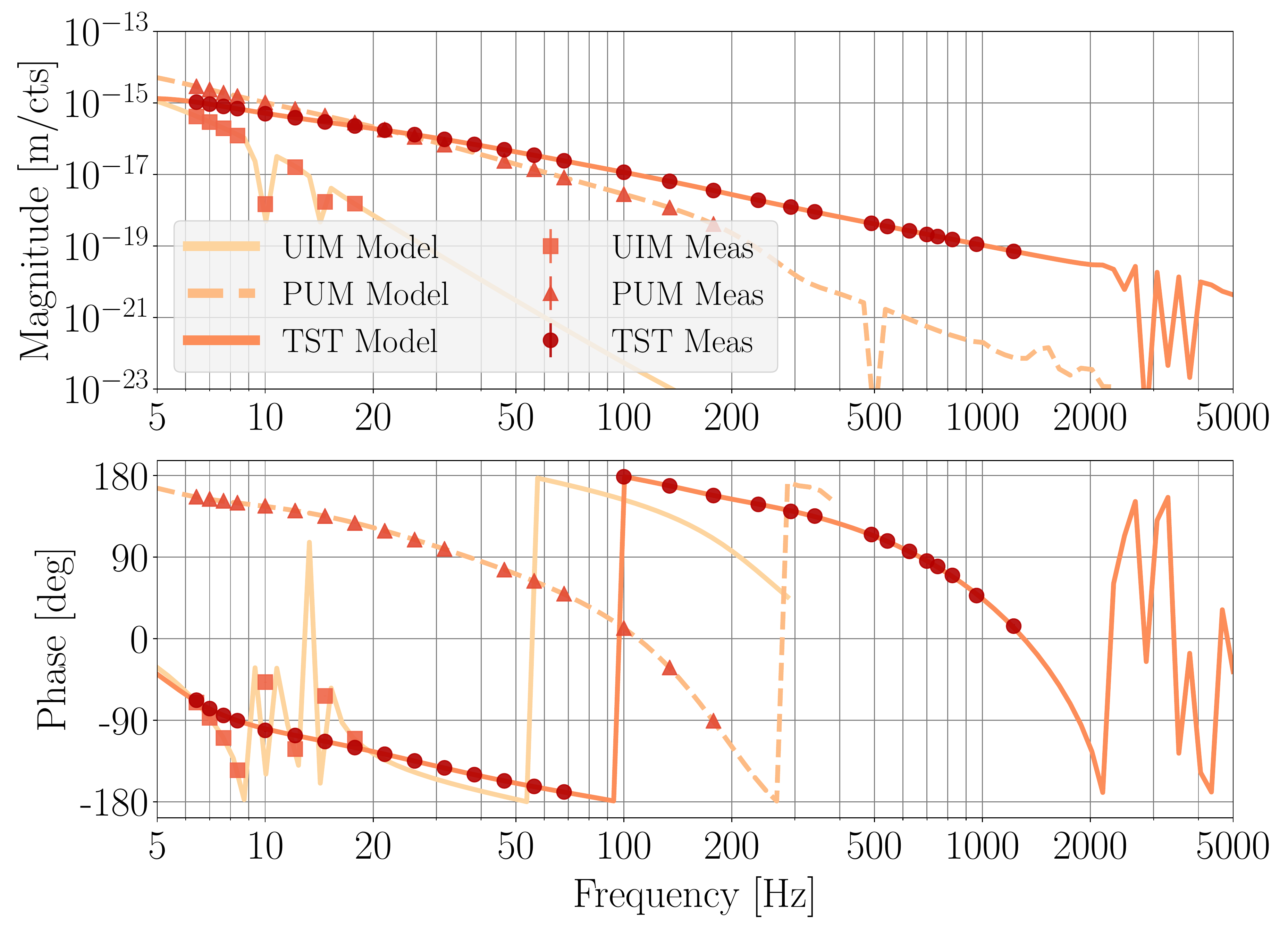}
\includegraphics[width=1.0\columnwidth] {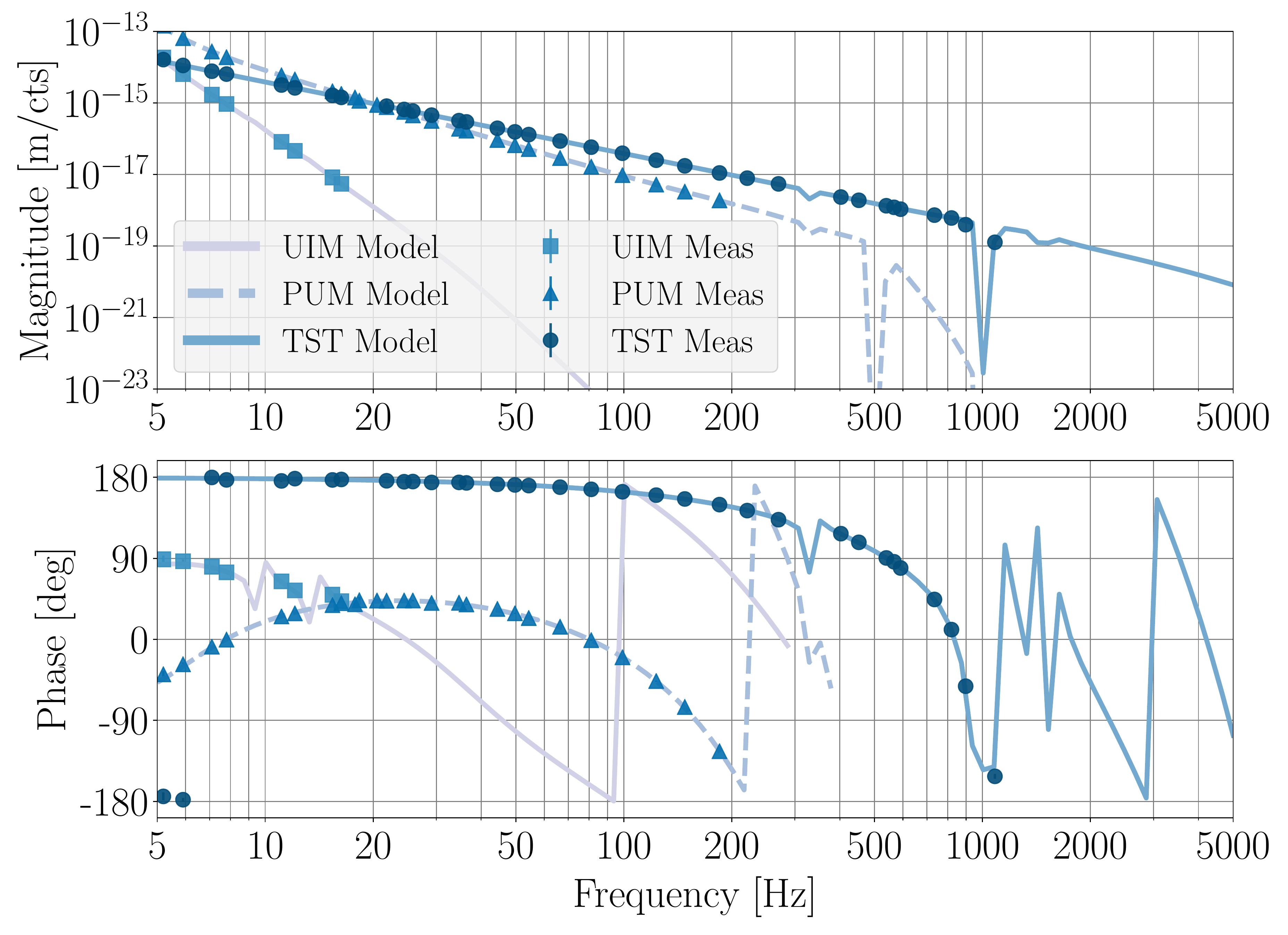}
\caption{Actuation stage measurements $[H_{i} \, A_{i}(f)]^{(\text{meas})}$ and models $[H_{i} \, A_{i}(f, \vec{\lambda}_{i})]^{(\text{model})}$.
Each index $i$ is one of the actuation stages $U$, $P$, or $T$.
The H1 actuation reference measurements from January 4th, 2017 are shown in the two left plots in red. 
The L1 actuation reference measurements from November 26th, 2016 are in the two right panels in blue.
The model parameters $\vec{\lambda}_{A}$ for $A_{i}(f, \vec{\lambda}_{i})$ have been found via the MCMC method.
The actuation strength magnitude is in units of meters per $d_{\text{ctrl}}$ count. 
Notches seen in the magnitude plot are purposefully placed to avoid ringing up suspension violin modes at specific frequencies.
Each stage's phase is sensible for frequencies at which each actuation stage dominates, but then rolls rapidly as it loses authority at high frequencies.
For this reason, the UIM and PUM stage phase plots are cut off at 300 Hz and 400 Hz respectively.
}
\label{fig:ActuationPlant}
\end{figure*}

\subsection{Measurements}
\label{subsec:Measurements}
In this section, we first describe our fundamental displacement reference, the radiation pressure actuator. 
Then we explore how that reference is used to measure the detector's response to DARM motion, or in other words, calibrate the detector.

The DARM model functions $C(f,t)$ and $A(f,t)$ are measured from swept sine transfer functions of the DARM control loop.
A swept sine transfer function is a collection of single frequency excitations applied in successive steps across the relevant frequency band of the detector.
The cross-correlation of actuator excitation against detector response during the excitation forms the transfer function.

The swept sine transfer functions are then manipulated to give transfer function measurements of each of the actuation stages and the sensing function.
Measurements of the detectors' DARM control loops require the detectors to be running at low-noise observation sensitivity.
Once a full suite of reference measurements is taken, the complete response of the detector to GWs can be estimated.

\subsubsection{Radiation Pressure Actuator}
\label{subsubsec:RadiationPressureActuator}

Two 1047 nm auxiliary laser systems known as photon calibrators (PCAL) displace each end test mass via radiation pressure \cite{aLIGOPCALPaper}.
The PCAL serves as a reference actuator on the test mass controlling the DARM loop.
The displacement of the test mass caused by the photon calibrator $x_T^{(PC)}$ is several orders of magnitude larger than the nominal displacement noise of the detector $\Delta L_{\text{free}}$ with integration time of 1 minute, which allows for high-precision characterization of the global control system and detector readout (See Figure \ref{fig:DARMControlLoop}).
Acousto-optic modulators (AOMs) are used to modulate the laser power incident on the test masses with arbitrary waveforms.
The power incident on the test masses is recorded via two photodiodes coupled to integrating spheres, one after the AOM before transmission onto the test mass, the other upon reflection off of the test mass. 
Each photodiode's readout is then digitally recast as a displacement, $x_{T}^{(PC)}$, which is the amount of PCAL-induced displacement contributing to $\Delta L_{\text{free}}$. 
The reflection photodiode was used for reference in all measurements described below.

The PCAL laser can introduce elastic deformations on the test mass, which can affect the calibration accuracy above 1 kHz.
Elastic deformation can be largely mitigated through the use of two beams symmetrically displaced from the center of the test mass \cite{aLIGOPCALPaper}.
The uncertainty budget does not include error from elastic deformation, assuming this effect is negligible up to 5 kHz.
The full suspension dynamics are incorporated into the transfer function from the PCAL power modulation to the test mass length modulation, giving an accurate frequency response at and below the suspension resonant frequency.

We are sometimes susceptible to systematic errors from PCAL ``clipping'' where the photon calibrator laser slightly misses the receiving photodiode, causing miscalibrations.
Fortunately, PCAL clipping systematic errors are quantifiable and included in the error budget.
The relative PCAL actuation strength correction factor, $H_{PCAL}(t)$, tracks the actuation strength of the PCAL over time.
$H_{PCAL}(t)$ has a value of 1 during times of no clipping, and a value less than 1 during times of clipping.
$H_{PCAL}(t)$ has a relative uncertainty of 0.79\% over all time.
This will affect our total calibration uncertainty budget directly in Section \ref{subsec:NumericalUncertaintyBudget}.
More on PCAL clipping is discussed in Section \ref{sec:Results}.
Further details of the PCAL and the composition of their uncertainty can be found in \cite{aLIGOPCALPaper}. 

Checks of gross systematic errors in the photon calibrator system have been performed using other, less precise displacement (or equivalent there-of) references and found agreement with the PCAL to within 10\% \cite{GW150914CalPaper}.

\subsubsection{Measurement Techniques}
\label{subsubsec:MeasurementTechniques}

To measure the PCAL to DARM transfer function, a known photon calibrator sine wave excitation $x_T^{(PC)}$ is applied to the detector while the DARM error signal $d_{\text{err}}$ is recorded.
This excitation is suppressed by the DARM control loop, forming the transfer function
\begin{align}
\label{xPCAL2DARMTF} \dfrac{d_{\text{err}}(f)}{x_T^{(PC)}(f)} = \dfrac{C(f)}{1 + G(f)}.
\end{align} 
The measurement suite is a collection of discrete sine waves swept over the frequency range $5$ Hz $ < f < 1$ kHz.
The closed loop suppression, $1/[1 +  G(f)]$, is then measured independently with the standard in-loop suspension actuators at the same frequencies as Equation \ref{xPCAL2DARMTF}. 
During times of clipping, we underestimate the excitation $x_T^{(PC)}$ by the relative actuation strength $H_{PCAL}(t)$, and must divide $x_T^{(PC)}$ by $H_{PCAL}(t)$ to correct for this.
The measured sensing function is then constructed as a function of frequency:
\begin{align}
\label{eq:SensingMeas} C^{(\text{meas})}(f) = H_{PCAL}(t) \, \left[1 + G(f)\right] \, \dfrac{d_{\text{err}}(f)}{x_T^{(PC)}(f)}.
\end{align}

Above 1 kHz, the photon calibrator's signal-to-noise ratio and actuation strength are low. 
In this region, the open loop gain $G(f)$ is negligible, so 
\begin{align}
\label{eq:HFSensingMeas} \dfrac{d_{\text{err}}(f)}{x_T^{(PC)}(f)} \approx \dfrac{C^{(\text{meas})}(f)}{H_{PCAL}(t)}, \qquad f > 1\,\text{kHz}
\end{align}
We obtain the sensing function at high frequency by performing a long-duration swept sine transfer function measurement. 
Each single frequency is driven for many hours, and the response is compensated for time dependence using only $\kappa_{C}(t)$. 

To measure the three actuation stages, similar swept sine excitations, $x_{i}(f)$, are applied to each stage at points upstream of the known distribution filters, $F_{i}(f)$, such that the detector readout measures
\begin{align}
\label{eq:xACT2DARMTF} \dfrac{d_{\text{err}}(f)}{x_i(f)} = \dfrac{H_{i} \, A_i(f) \, C(f)}{1 + G(f)}
\end{align}
where the index $i$ indicates either the upper intermediate $U$, penultimate $P$, or test mass $T$ stages.
These excitations are then compared to an excitation from the photon calibrator to isolate each actuation plant, as in Eq. \ref{xPCAL2DARMTF}, to form
\begin{align}
\label{eq:ActuationMeas} [H_{i} \, A_i(f) ]^{(\text{meas})} =  \dfrac{1}{H_{PCAL}(t)} \, \dfrac{x_T^{(PC)}(f)}{d_{\text{err}}(f)} \, \dfrac{d_{\text{err}}(f)}{x_i(f)}.
\end{align}

The relative magnitude uncertainty and absolute phase uncertainty in a transfer function swept sine measurement point is calculated by Bendat and Piersol \cite{BendatPiersolCoherenceUncertainty}:
\begin{align}
\label{eq:CoherenceUncertainty} \sigma^{(\text{meas})}(f) = \sqrt{ \dfrac{1 - \gamma^2(f)}{2 \, N_{\text{avg}} \, \gamma^2(f)} }
\end{align}
where $\gamma^{2}(f)$ is the coherence between excitation and readout, and $N_{\text{avg}}$ is the number of averages at each excitation frequency point.

To capture the time dependence of the calibration during a run, ``calibration lines'' are applied to the detectors during all observation times.  
A calibration line is a single-frequency excitation applied to the detector via the photon calibrator and suspension actuators.
Using four calibration lines, we are able to capture changes in the detector calibration and partially correct for them in real time. 

The calibration lines' response to the applied excitation is recorded in the detector readout $d_{\text{err}}$.
These transfer functions are recast into each time dependent parameter, $\kappa_{T}$, $\kappa_{PU}$, $\kappa_{C}$, and $f_{CC}$.
The calibration lines are driven with high signal-to-noise ratio such that the time-dependent parameter uncertainties are small relative to the parameter values.
The calculation of the time-dependent parameters from calibration lines is derived in \cite{CALTimeDependence}.

\begin{figure}
\includegraphics[width=0.48\textwidth]{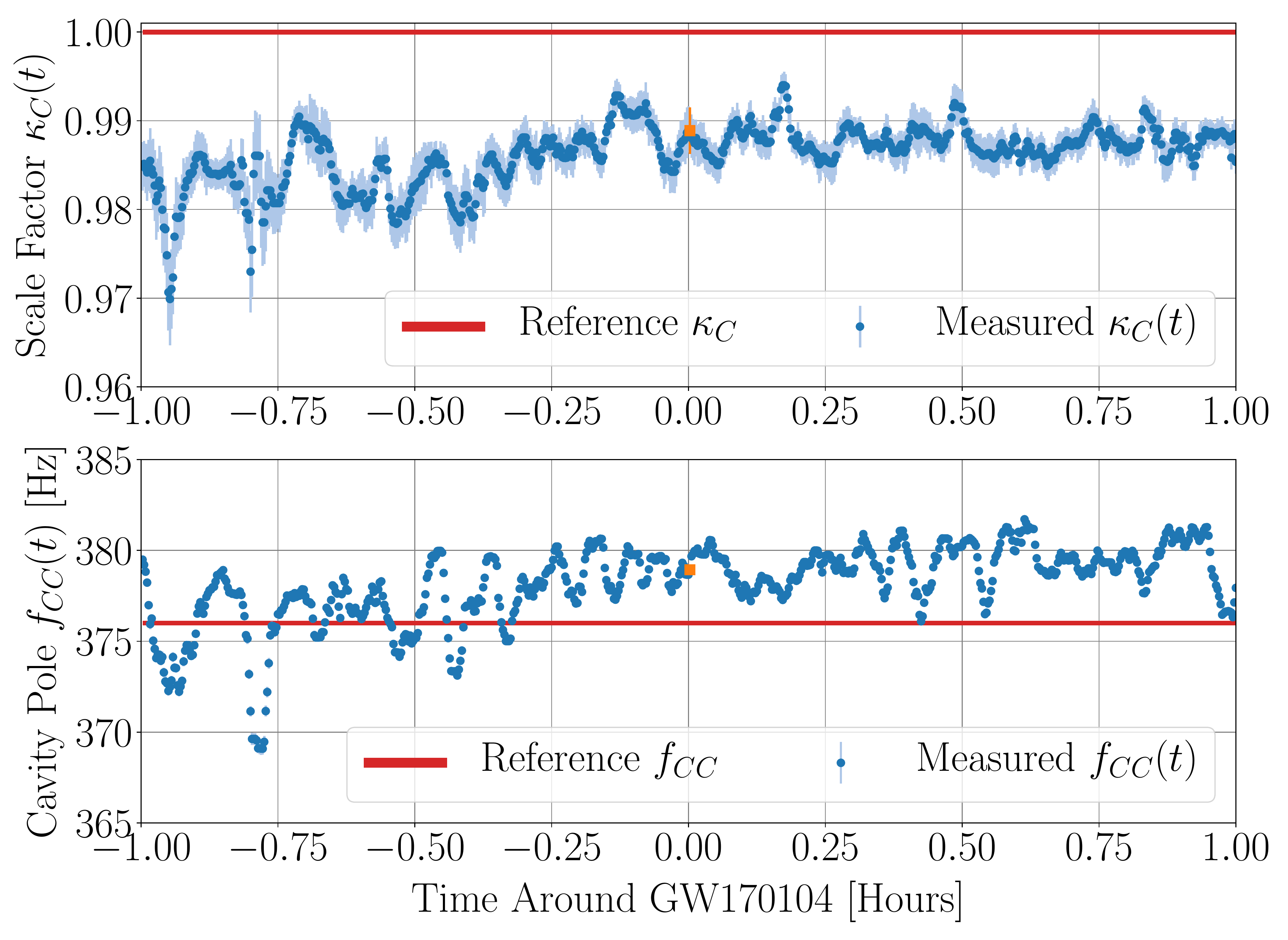}
\caption{Time Dependent Parameters in L1 around GW170104. \\
The time series of the optical gain scale factor $\kappa_{C}(t)$ and coupled cavity pole $f_{CC}(t)$ in L1 for the two hours surrounding GW170104.
The reference values are plotted as a red line, while the data is blue dots with uncertainty bars.  
The values at the time of GW170104 are plotted in orange.
Changes in $\kappa_{C}(t)$ are updated in the reference sensing model $C^{(\text{model})}(f,t)$.
Changes in $f_{CC}(t)$ are not reflected in the reference sensing model.
This represents a systematic error in the calibration pipeline strain data $h(t)$.
}
\label{fig:KappaUncertainty}
\end{figure}

The statistical uncertainty in a time-dependent parameter $\sigma_{\kappa_{i}}(t)$, at any given time, $t$, is derived from the measured coherence of the calibration lines used to form them (see Equation \ref{eq:CoherenceUncertainty}, propagated as in \cite{CALTimeDependence}).
That uncertainty is then used to form a distribution, with mean and standard deviation of $\kappa_{i}(t)$ and $\sigma_{\kappa_{i}}(t)$.
The posteriors of this distribution are sampled and propagated through to the total uncertainty for that given time, as described below in Section \ref{subsec:NumericalUncertaintyBudget}. 
See Figure \ref{fig:KappaUncertainty} for an example result from this process for L1's $\kappa_{C}(t)$ at the time of GW170104.

\begin{table*}

\begin{center}
\caption{Sensing Parameters \\
H1 (left) and L1 (right) sensing function model parameters $\vec{\lambda}_{C}$ MCMC fit values  and uncertainties.  
The fits were performed on H1's January 4th, 2017 reference measurement and L1's November 26th, 2016 reference measurement.
The model corresponding to these parameters can be seen in Figure \ref{fig:SensingPlant}.
The corner plot showing the MCMC results from the H1 reference measurement is shown in Figure \ref{fig:SensingCornerplot}. \\
} 
\label{tab:SensingTable}
\begin{minipage}{0.48\linewidth}
\bgroup
\def\arraystretch{1.5}
\begin{tabular}{ c c c c }
H1 Parameters                 & Variable   & Value${}_{-1\sigma}^{+1\sigma}$ & Units \\ 
\hline
Optical Gain              & $H_C$     & $3.834_{-0.003}^{+0.003}$ & mA/pm \\  
Coupled Cavity Pole & $f_{CC}$  & $360_{-2}^{+2}$  & Hz \\
Time Delay                & $\delta \tau_{C}$      & $0.6_{-1.3}^{+1.3}$  & $\mu$s \\
Optical Spring Frequency      & $f_S$      & $6.87_{-0.03}^{+0.03}$  & Hz\\
Optical Spring Inverse Q       & $Q_S^{-1}$  & $0.034_{-0.004}^{+0.004}$  & none \\
\hline
\end{tabular}
\egroup
\\[5pt]
\end{minipage}
\quad
\begin{minipage}{0.48\linewidth}
\label{tab:LLOSensingTable}
\bgroup
\def\arraystretch{1.5}
\begin{tabular}{ c c c c }
L1 Parameters                 & Variable   & Value${}_{-1\sigma}^{+1\sigma}$ & Units \\ 
\hline
Optical Gain              & $H_C$           & $3.288_{-0.007}^{+0.007}$ & mA/pm \\  
Coupled Cavity Pole & $f_{CC}$  & $369.5_{-0.9}^{+1.0}$  & Hz \\
Time Delay                & $\delta \tau_{C}$      & $-0.84_{-0.13}^{+0.13}$  & $\mu$s \\
Optical Spring Frequency      & $f_S$      & $2.6_{-0.2}^{+0.2}$  & Hz\\
Optical Spring Inverse Q       & $Q_S^{-1}$  & $0.005_{-0.004}^{+0.009}$  & none \\
\hline
\end{tabular}
\egroup
\\[5pt]
\end{minipage}

\end{center}
\end{table*}

\begin{table*}

\begin{center}
\caption{Actuation Parameters \\
H1 (left) and L1 (right) actuation function model parameters $\vec{\lambda}_{A}$ MCMC fit values and uncertainties.  
The fits were performed on H1's January 4th, 2017 reference measurements and L1's November 26th, 2016 reference measurement.
The models corresponding to these parameters can be see in Figure \ref{fig:ActuationPlant}.
To get from Newtons/count units in this table to meters/count in Figure \ref{fig:ActuationPlant}, we multiply by the suspension models which have units of meters/Newton and are known to negligible uncertainty.
} 
\label{tab:ActuationTable}
\begin{minipage}{0.48\linewidth}
\bgroup
\def\arraystretch{1.5}
\begin{tabular}{ c c c c }
H1 Parameters                 & Variable   & Value${}_{-1\sigma}^{+1\sigma}$ & Units \\ 
\hline
Upper Intermediate Gain    & $H_{U}$      & $8.205_{-0.004}^{+0.004} \times 10^{-8}$ & N/cts \\  
Upper Intermediate Delay  & $\delta \tau_{U}$  & $57_{-46}^{+45}$  & $\mu$s \\
Penultimate Gain                & $H_{P}$       & $6.768_{-0.002}^{+0.002} \times 10^{-10}$  & N/cts \\
Penultimate Delay              & $\delta \tau_{P}$   & $0.4_{-0.6}^{+0.6}$  & $\mu$s\\ 
Test Mass Gain                   &$H_{T}$       & $4.3573_{-0.0008}^{+0.0008} \times 10^{-12}$  & N/cts\\
Test Mass Delay                 & $\delta \tau_{T}$  & $2.8_{-0.4}^{+0.4}$  & $\mu$s \\
\hline
\end{tabular}
\egroup
\\[5pt]
\end{minipage}
\quad
\begin{minipage}{0.48\linewidth}
\label{tab:LLOActuationTable}
\bgroup
\def\arraystretch{1.5}
\begin{tabular}{ c c c c }
L1 Parameters                 & Variable   & Value${}_{-1\sigma}^{+1\sigma}$ & Units \\ 
\hline
Upper Intermediate Gain    & $H_{U}$      & $7.24_{-0.03}^{+0.03} \times 10^{-8}$ & N/cts \\  
Upper Intermediate Delay  & $\delta\tau_{U}$  & $102_{-56}^{+56}$  & $\mu$s \\
Penultimate Gain                & $H_{P}$       & $6.41_{-0.02}^{+0.02} \times 10^{-10}$  & N/cts \\
Penultimate Delay              & $\delta\tau_{P}$    & $-8.7_{-6.1}^{+6.2}$  & $\mu$s\\ 
Test Mass Gain                   &$H_{T}$        & $2.513_{-0.004}^{+0.004} \times 10^{-12}$  & N/cts\\
Test Mass Delay                 & $\delta\tau_{T}$   & $-4.5_{-1.4}^{+1.4}$  & $\mu$s \\
\hline
\end{tabular}
\egroup
\\[5pt]
\end{minipage}
\end{center}

\end{table*}

\section{Uncertainty and Error Estimation}
\label{sec:Method}
Our uncertainty budget is numerically evaluated by producing a large number of realizations of the response function.
To do this, we first estimate the DARM model parameters using a Markov Chain Monte Carlo (MCMC) method.
Next, we stack all measurement residuals and estimate any deviations from the model using a Gaussian process regression.
Then, we sample our MCMC and regression results to form ten thousand resultant response functions.
These response functions stacked form the calibration error and uncertainty budget.

\subsection{DARM Model Parameter Estimation}
\label{subsec:MCMC}

First, a measurement $\vec{d} = $ $C^{(\text{meas})}(f)$ or $A^{(\text{meas})}(f)$ is obtained as described in Section \ref{subsec:Measurements}.
Next, the models $\vec{M} = $ $C^{(\text{model})}(f,t,\vec{\lambda}_{C})$ or $A^{(\text{model})}(f,t,\vec{\lambda}_{A})$ are fit to the measurement by varying the model parameters $\vec{\lambda} = $ $\vec{\lambda}_{C}$ or $\vec{\lambda}_{A}$ via a Markov Chain Monte Carlo (MCMC).

An MCMC algorithm can quickly approximate the posterior probability distributions on the values of the model parameters given a log likelihood function and assumed prior distribution.
The log likelihood, $\log \mathcal{L}(\vec{M} \, | \, \vec{\lambda}, \vec{d})$, is a simple least squares comparison between the model values $\vec{M}(\vec{\lambda})$ (where $\vec{M}$ represents $C^{\text{(model)}}$ or $A^{\text{(model)}}$) given model parameters $\vec{\lambda}$ (namely $\vec{\lambda}_{C}$ or $\vec{\lambda}_{A}$) and measurement data $\vec{d}$ (as described in Section \ref{subsec:Measurements}).
All initial parameter estimates in $\vec{\lambda}_{C}$ and $\vec{\lambda}_{A}$ were assumed to have flat prior distributions.
The maximum a posteriori (MAP) values of the posterior distributions are taken as the best fit values.
The ensemble of MCMC distributions are saved to be sampled for the total uncertainty budget in subsection \ref{subsec:NumericalUncertaintyBudget}.

\begin{figure}
\includegraphics[width=0.48\textwidth]{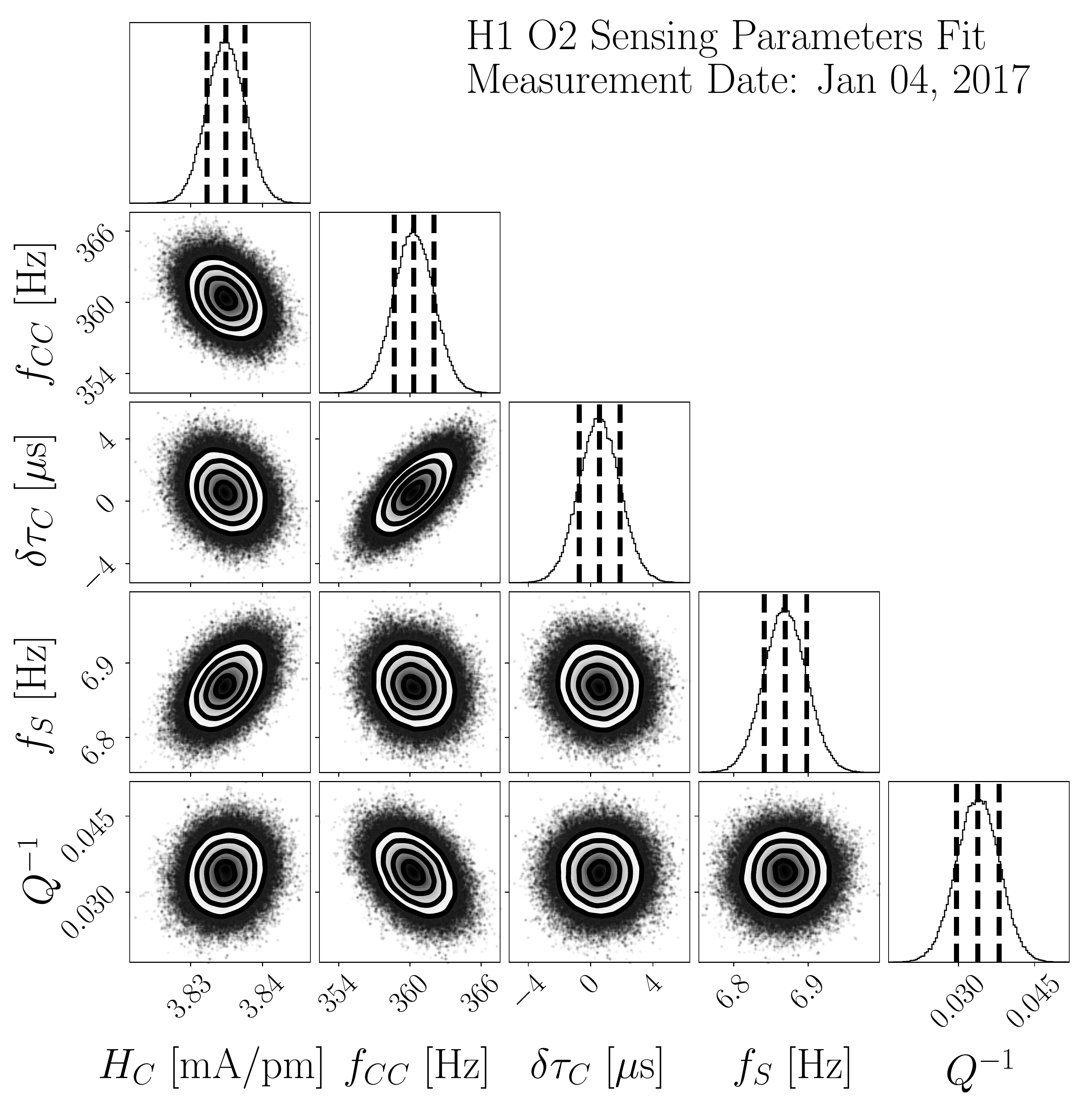}
\caption{Posterior distribution on the H1 sensing parameters $\vec{\lambda}_{C}$.  \\
Each column represents one of the five sensing parameters: optical gain $H_C$, coupled cavity pole $f_{CC}$, time delay correction $\delta \tau_{C}$, optical spring $f_S$, and optical spring inverse quality factor $Q_S^{-1}$.
Each point represents a sample in five dimensional parameter space.
The diagonal plots represent the variance on each parameter, while the off-diagonal plots show the covariance of each parameter with another.
The dashed vertical lines on the diagonal plots represent the median and $1\sigma$ values for each parameter.
}
\label{fig:SensingCornerplot}
\end{figure}

The MCMC posteriors are found for both detector's frequency dependent models: $C^{(\text{model})}(f,t,\vec{\lambda}_{C})$ and $A_{i}^{(\text{model})}(f,t,\vec{\lambda}_{i})$.
The best fit values are reported in Tables \ref{tab:SensingTable} and \ref{tab:ActuationTable}.
The plots of the model fits can be seen in Figures \ref{fig:SensingPlant} and \ref{fig:ActuationPlant}.
The one- and two-dimensional posterior distributions for the H1 sensing model parameters $\vec{\lambda}_{C}$ are shown in Figure \ref{fig:SensingCornerplot}.
The MCMCs were performed using the python \texttt{emcee} toolbox \cite{matplotlib, emcee}.
The plot was produced with the \texttt{corner} python plotting package \cite{22}.

\subsection{Quantifying Frequency Dependent Error and Uncertainty}
\label{subsec:GP}
Throughout observing runs, collections of detector measurements are taken regularly. 
Every measurement taken is run through the MCMC method as detailed in subsection \ref{subsec:MCMC}.
The measurement is then divided by its best fit DARM model to produce a residual, as seen in Equation \ref{eq:TFSystErrors}.

All of the residuals are gathered together into a collection of all measurements taken over the observing run.
These residuals have all known systematic errors removed, but still contain information about unknown systematic errors.
We create a distribution of functions that could describe this residual systematic error, then we incorporate this distribution into the calibration uncertainty budget.  
To accomplish this, we use a Gaussian process regression \cite{Rasmussen2006, sklearn}.

A Gaussian process is a method of producing distributions over random functions.  
The Gaussian process regression takes in data and a user-defined covariance matrix, tunes the covariance matrix hyperparameters to fit the given data, and outputs a posterior of potential function fits to the data.
This allows an uncertainty budget to be produced for arbitrary frequencies, creating a continuous posterior distribution from discrete data.

From the resulting posterior distribution, we can extract a most probable fit function, known as the mean function.
The mean function becomes the systematic error $\delta C^{GP}(f)$ and $\delta A_{i}^{GP}(f)$ in Equations \ref{eq:SensingError} and \ref{eq:ActuationError}.
We can also draw frequency dependent uncertainties $\sigma_{\delta C}^{GP}$ and $\sigma_{\delta A_{i}}^{GP}$ on the systematic error.
Posteriors representing $\sigma_{\delta C}^{GP}$ and $\sigma_{\delta A_{i}}^{GP}$ will be sampled for the total uncertainty budget in subsection \ref{subsec:NumericalUncertaintyBudget}.

Our Gaussian process regression trains on the residual data with the following covariance kernel 
\begin{align}
\label{eq:CovarianceKernel} \nonumber k\left(\log(f), \log(f')\right) &= \gamma_1^2 + \log(f) \cdot \log(f') \\
\nonumber &+ \left( \gamma_2^2 + \log(f) \cdot \log(f')\right)^2 \\
&+ \gamma_3^2\, \exp\left(-\dfrac{\left(\log(f) - \log(f')\right)^2}{2 \ell^2}\right)
\end{align}
where $\left\{\gamma_1, \gamma_2, \gamma_3, \ell\right\}$ are the hyperparameters of the covariance kernel.  
The hyperparameters are tuned by the Gaussian process via gradient descent to best match the training data.
This kernel assumes the detector plants' systematic error should be characterized in the log frequency domain, and that the error is relatively smooth and can be captured by a squared exponential and quadratic kernel.

\begin{figure}
\includegraphics[width=0.48\textwidth]{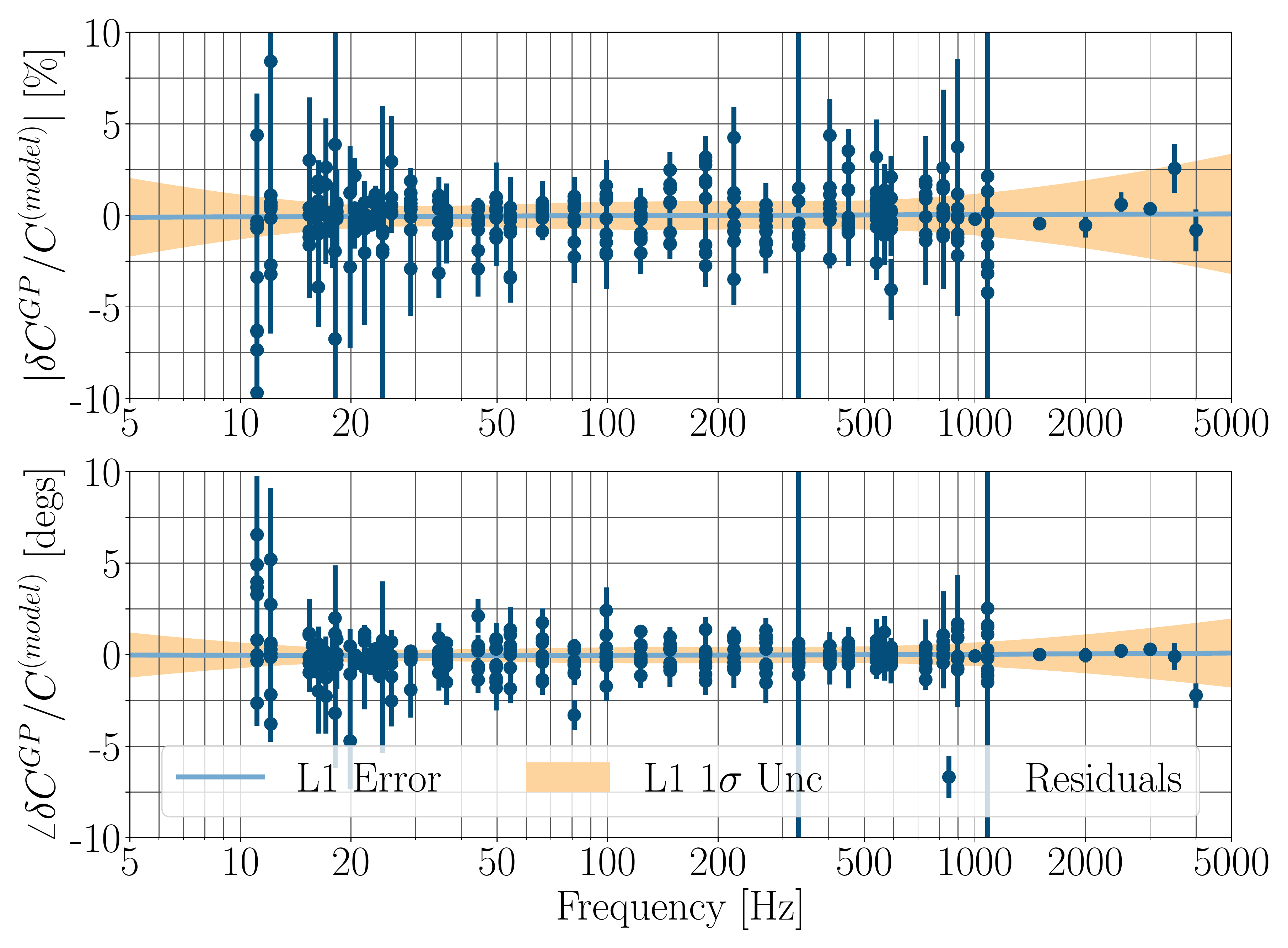}
\caption{Gaussian process regression of L1's sensing systematic error $\delta C^{GP}(f)$.  
The dark blue points are all the sensing measurement residuals, $\delta C/C^{(\text{model})}(f, t, \vec{\lambda}_{C})$, taken over the entire observation run.
This includes the residuals from the L1 reference measurement in the far right plots of Figure \ref{fig:SensingPlant}.
The light blue line is the mean function representing systematic error.
The light orange envelope is the $1\sigma$ uncertainty on the systematic error.
}
\label{fig:GaussianProcessRegression}
\end{figure}

An example collection of measurement residuals for the L1 detector's sensing function and the resulting Gaussian process regression is shown in Figure \ref{fig:GaussianProcessRegression}.
Here we show the same data from Figure \ref{fig:SensingPlant}, but with additional measurements from the entire observation run. 

\subsection{Total Calibration Uncertainty Budget}
\label{subsec:NumericalUncertaintyBudget}

The total calibration uncertainty budget for any given time is contructed from many sampled response functions $R(f, t)$.
Each sample response function is constructed by sampling from the posteriors of the response function components.
The response function components are:
\begin{enumerate}
  \item The sensing DARM model parameters: \\
  \centerline{$\vec{\lambda}_{C} = \left\{H_C, f_{CC}, \delta \tau_{C}, f_S, Q_S^{-1}\right\}$}
  \item The actuation DARM model parameters: \\
  \centerline{$\vec{\lambda}_{A} = \left\{H_{U}, \delta\tau_{U}, H_{P}, \delta\tau_{P}, H_{T}, \delta\tau_{T}\right\}$}
  \item The sensing Gaussian process systematic error: \\
  \centerline{$\delta C^{GP}(f)$}
  \item The actuation Gaussian process systematic errors: \\
  \centerline{$\delta A_{U}^{GP}(f)$, $\delta A_{P}^{GP}(f)$, $\delta A_{T}^{GP}(f)$}
  \item The time dependent parameters: \\
  \centerline{$\kappa_{T}(t)$, $\kappa_{PU}(t)$, $\kappa_{C}(t)$, $f_{CC}(t)$}
  \item The photon calibrator radiation pressure strength: \\
  \centerline{$H_{PCAL}(t)$}
\end{enumerate}
Each of these components to the response have had posterior distributions constructed previously:
(1) and (2) from the MCMC ensemble results on the reference measurements, (3) and (4) from the Gaussian process regressions on the residuals to incorporate unknown systematic errors, (5) from the calibration line measurements and coherence, and (6) from the 0.79\% uncertainty in $H_{PCAL}(t)$ from the photon calibrator paper \cite{aLIGOPCALPaper}.

Ten thousand samples are drawn from each of these posterior distributions and combined into ten thousand response function samples according to Equation \ref{eq:ResponseDef}.
Each of these response functions is then divided by the nominal response function, $R^{(\text{model})}(f, t)$, which is constructed from the sensing model $C^{(\text{model})}(f, t, \vec{\lambda}_{C})$ and actuation model $A^{(\text{model})}(f, t, \vec{\lambda}_{A})$.
This gives ten thousand relative response functions, each of which is plotted in Figure \ref{fig:O2ResultsPlot}.  
The median of this relative response function distribution constitutes the overall systematic error, and the 68th percentile upper and lower contours are the statistical uncertainty, both a function of frequency.

Figure \ref{fig:O2ResultsPlot} shows the calibration uncertainty at the time of the most recent detection, GW170104. 
Table \ref{tab:GW170104ExtremeUncertainty} reports the ``extreme uncertainty'' for calibration between 20-1024 Hz during GW170104.  
Extreme uncertainty refers to the maximum and minimum of the systematic error $\pm 1\sigma$ uncertainty within a certain frequency band.
This quantity is useful for searches requiring single number calibration uncertainty values, and ignore calibration systematic errors or frequency-dependent calibration uncertainty.

\subsection{Calibration Uncertainty for Entire Observing Runs}
Calibration error and uncertainty evolves over observing runs, affecting the results of continuous and stochastic gravitational wave searches \cite{O1CW2017a, O1CW2017b, O1Stoch2017a, O1Stoch2017b}.
To assess the uncertainty of the detectors throughout an observing run, a total calibration uncertainty budget is made for every hour of observing data. 

Collapsing the uncertainty budgets along the time axis, the 68th, 95th, and 99th percentile ($1\sigma$, $2\sigma$ and $3\sigma$) limits are reported.
The entire run's calibration error and uncertainty is often reduced to a single statement such as ``over the course of an observing run, the $1\sigma$ uncertainty is no larger than XX \% in magnitude and YY degrees in phase.'' 
To do so, the extreme uncertainty is taken in magnitude (XX\%) and phase (YY degrees) using the 68th percentile contour over the relevant frequency band.

\section{Results}
\label{sec:Results}
\begin{figure*}
\includegraphics[width=0.48\textwidth]{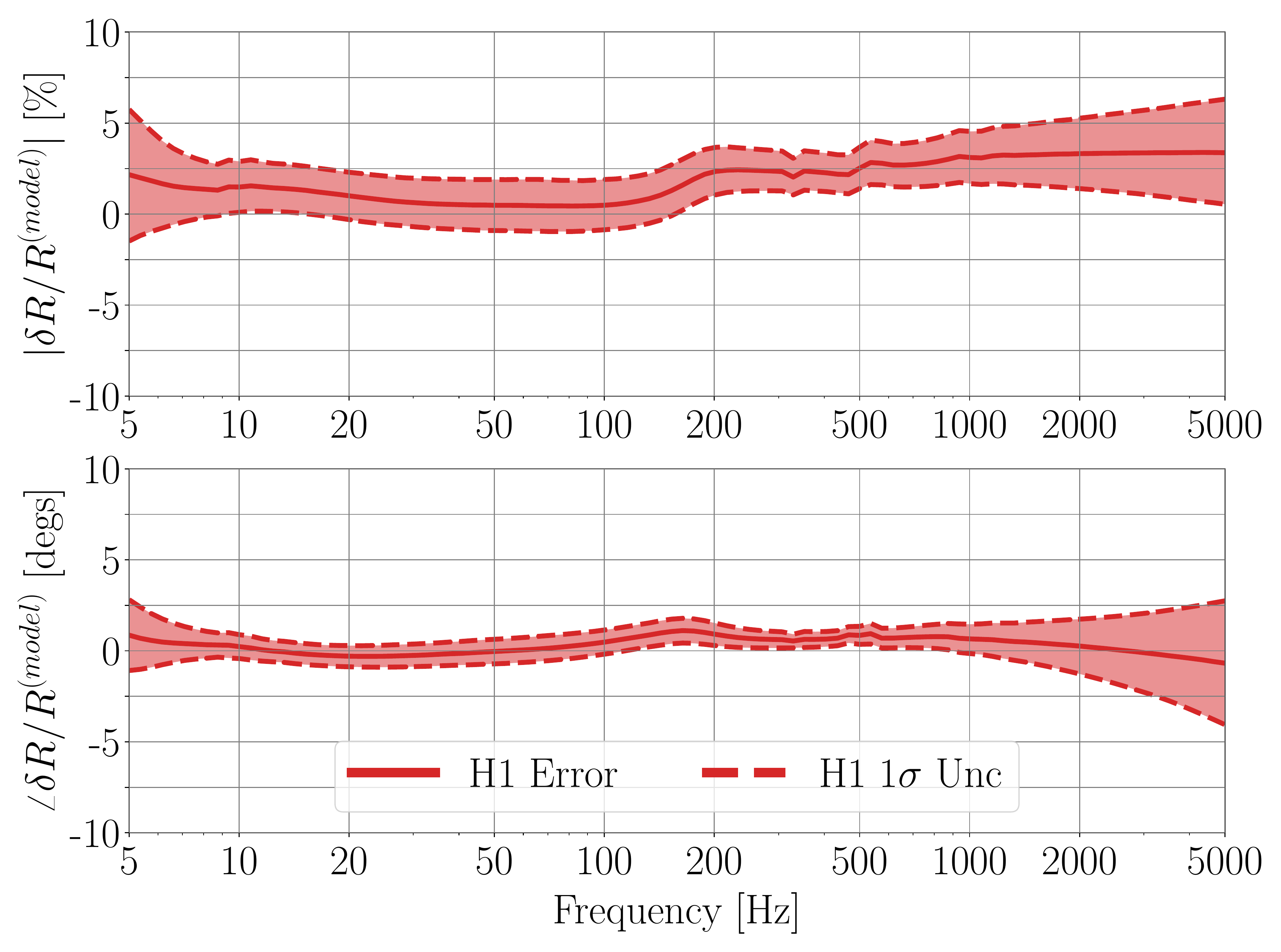}
\includegraphics[width=0.48\textwidth]{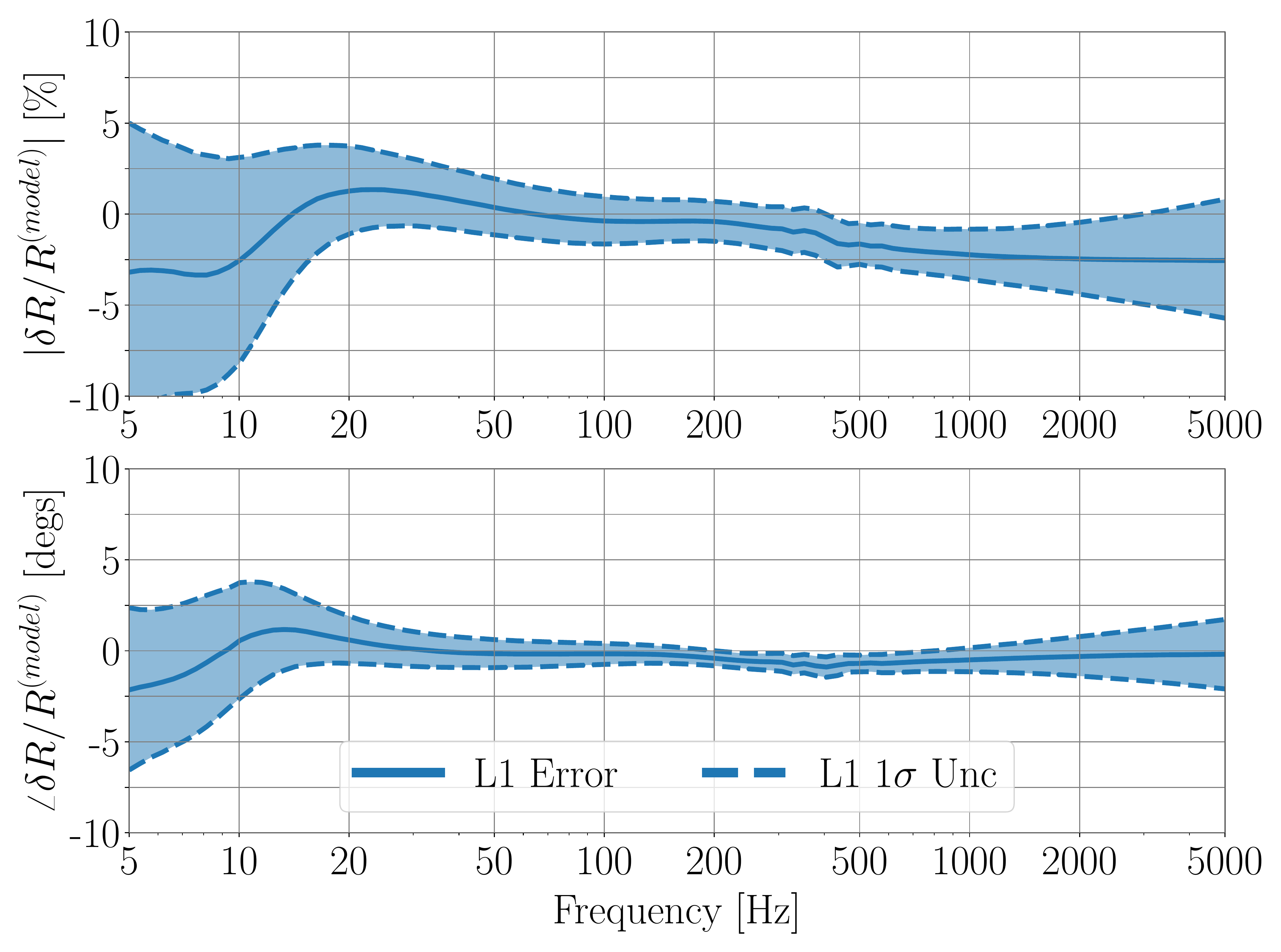}
\caption{Total Calibration Error and Uncertainty Budget at the time of GW170104.\\
The uncertainty in the calibrated response function for the H1 detector is on the left, and for L1 is on the right.
The $y$ axis is relative response error $\delta R / R^{(\text{model})}$ and uncertainty $\sigma_{R} / R^{(\text{model})}$, with magnitude on top and phase on the bottom.  
The solid line is the median relative response, interpreted as the frequency dependent systematic error on the model response $R^{(\text{model})}$.
The dashed lines represent the $1\sigma$ uncertainty on this error.  
Stacking ten thousand drawn response function samples produces the numerical uncertainty budget shown here.
The extreme $1\sigma$ uncertainties are presented in Table \ref{tab:GW170104ExtremeUncertainty}. 
}
\label{fig:O2ResultsPlot}
\end{figure*}

The final calibration uncertainty budget for GW170104 is shown in Figure \ref{fig:O2ResultsPlot}.
The ``extreme uncertainties'', or the maximum and minimum of error $\pm 1\sigma$ uncertainty, are reported in Table \ref{tab:GW170104ExtremeUncertainty}.

The previous uncertainty quantification method from \cite{GW150914CalPaper} conservatively reported 10\% and 10 degrees uncertainties for GW150914 and the calibration uncertainties for all three O1 events in Table III in \cite{O1BBHPaper}.
The uncertainty quantification method used for GW170104 was repeated on the O1 events.
These results are reported in Appendix \ref{sec:appendixA}, with plots of the uncertainty budgets for GW150914, LVT151012, and GW151226 in Figure \ref{fig:O1ResultsPlot} and extreme uncertainties reported in Table \ref{tab:O1EventsExtremeUncertainty}.

Systematic errors are known discrepancies between the detector model and measurement.
At low frequency, the systematic error is dominated by the Gaussian process regression on the actuation function residuals.
At high frequency, fluctuations in the coupled cavity pole $f_{CC}(t)$, which are not corrected for in the calibration procedure, dominate the error budget.

Uncertainty everywhere is dominated by the Gaussian process regression on both functions.  
The uncertainty from the MCMC parameter fits on $\vec{\lambda}_{C}$ and $\vec{\lambda}_{A}$, and the uncertainty in the time dependent parameters $\kappa_{T}(t)$, $\kappa_{PU}(t)$, $\kappa_{C}(t)$, and $f_{CC}(t)$ tend to be about an order of magnitude smaller than the Gaussian process regression results.
The 0.79\% uncertainty in the photon calibration strength $H_{PCAL}(t)$ contributes only to magnitude uncertainty.

\begin{figure*}
\includegraphics[width=0.48\textwidth]{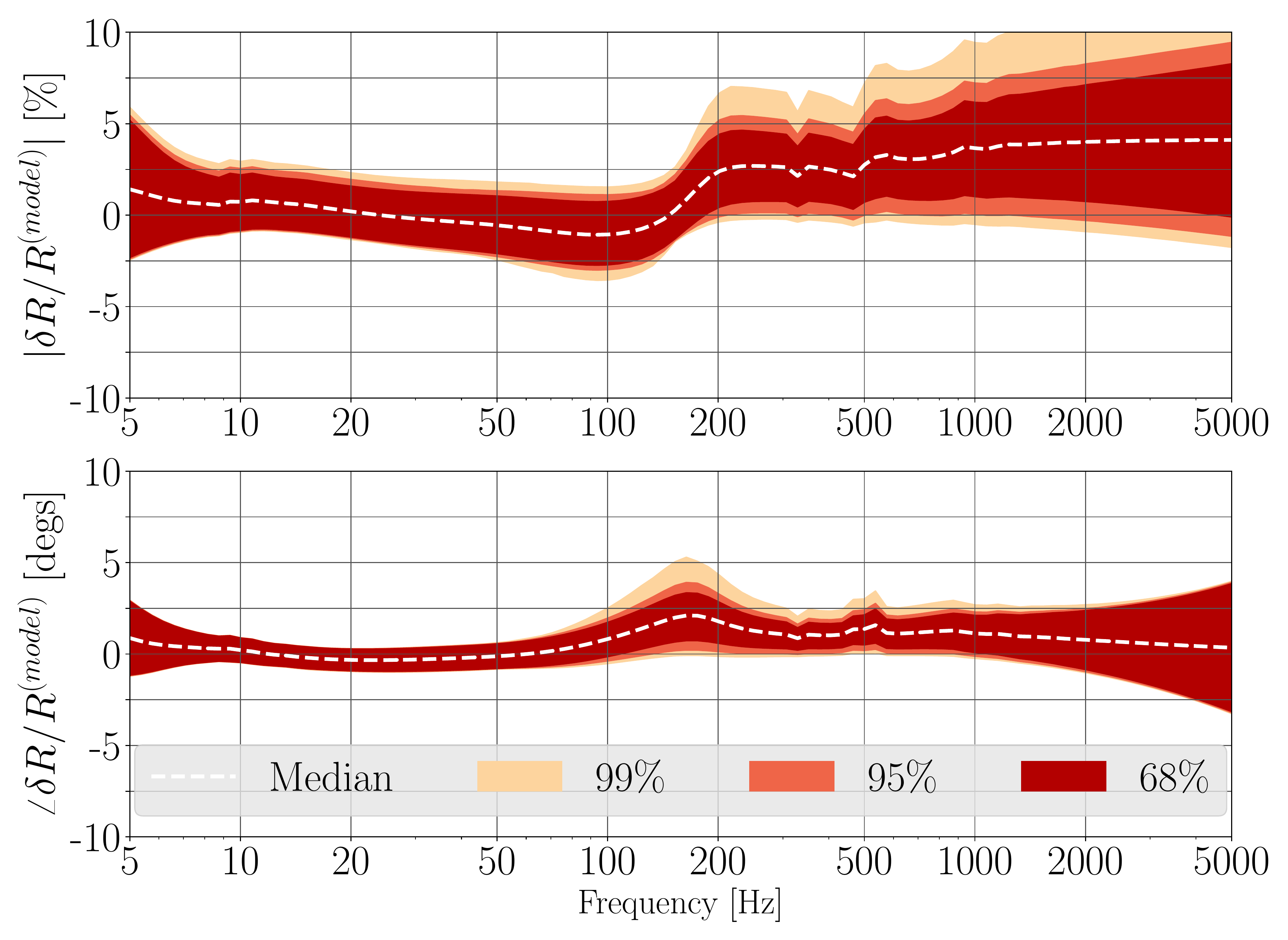}
\includegraphics[width=0.48\textwidth]{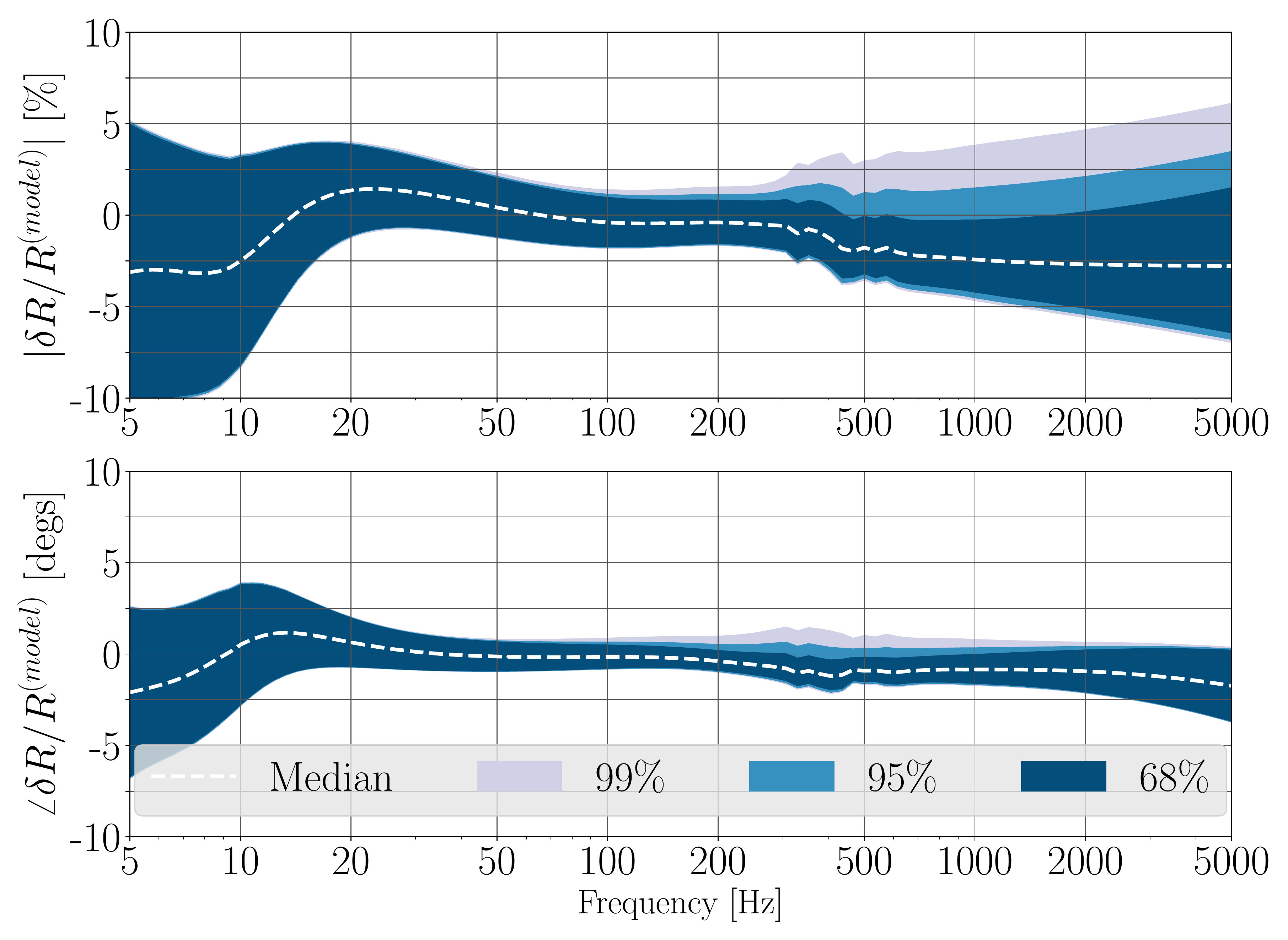}
\caption{Total Calibration Uncertainty Percentiles for Observing Run Two.\\
The percentiles are created for all of O2 data from November 30, 2016 to August 25th, 2017.
H1's uncertainty is on the left, and L1's is on the right.
The $y$ axis is relative response $\delta R/R^{(\text{model})}$ magnitude (top) or phase (bottom), stacked for all times in the observing run.
The dashed white line is the median relative response, while the colors represent the $1\sigma$ calibration uncertainty for 68\%, 95\%, and 99\% of the run's time.
The largest changes in the calibration at H1 were due to clipping of the photon calibrator laser misreporting the strength of our response.
The largest calibration changes at L1 were due to fluctuations in the coupled cavity pole, which changes in time but is not yet corrected for in our calibrated data.
}
\label{fig:O2PercentilePlots}
\end{figure*}

The uncertainty and error for O2 strain data from November 19 through June 19 is shown in Figure \ref{fig:O2PercentilePlots}.
This percentile plot was created by taking all observing time, producing an uncertainty budget for each hour, then compiling each budget into the percentiles shown.
Overall, the detector calibration is stable over time.
This consistency is largely due to the correction of the scale factors $\kappa_{T}(t)$, $\kappa_{PU}(t)$, and $\kappa_{C}(t)$ in the calibration pipeline models.
Uncorrected systematic errors in the cavity pole $f_{CC}(t)$ are particularly visible at L1 at high frequency.

During some parts of the second observing run, we have found that the reflection photodetector of the PCAL system at the H1 detector had suffered from clipping.
Clipping means that the PCAL laser light incident on the photodetector was slightly off, giving a false low reading of how much power the PCAL was outputing.
This means any measurement taken using the reflection photodiode as reference had a systematic error in scale.
This includes the scale of any continuously measured time-dependent model parameters which are applied as correction factors for the estimated detector output, $h(t)$.
We have quantified this systematic error using the same system's transmission photodiode, and included it as systematic error in the overall response.
The systematic error was on the order of a few percent, and can be seen reflected in the upper percentiles of the H1 uncertainty in Figure \ref{fig:O2PercentilePlots}.

\begin{table}

\begin{center}
\caption{GW170104 Extreme Uncertainty \\
Below are the extreme calibration uncertainty values for H1 and L1 at the time of GW170104 in the 20-1024 Hz frequency range.
``Extreme uncertainty'' refers to the maximum and mininum of error $\pm 1\sigma$ uncertainty.
The plots informing this table can be seen at Figure \ref{fig:O2ResultsPlot}
} 
\label{tab:GW170104ExtremeUncertainty}
\bgroup
\def\arraystretch{1.5}
\begin{tabular}{ c c c }
GW170104 Uncertainty & H1 & L1 \\ 
\hline
$+1\sigma$ Magnitude [\%]    & 4.6 \%    &  3.7 \% \\
$-1\sigma$ Magnitude [\%]     & -1.0 \%  & -3.7 \% \\
$+1\sigma$ Phase [degrees]  & 1.8$^{\circ}$     &  1.9$^{\circ}$  \\
$-1\sigma$ Phase [degrees]   & -0.9$^{\circ}$    & -1.4$^{\circ}$ \\
\hline
\end{tabular}
\egroup
\\[5pt]
\end{center}
\end{table}

\section{Future Work}
\label{sec:FutureWork}

There is much to be done to build upon this work.
First, we will make use of calibration lines to track the detuning spring frequency $f_s$ and $Q$ values in real time. 
This will ensure the sensing plant is not severely detuned, or changing rapidly during detector operation.
Second, we will employ time domain filters capable of correction for frequency-dependent changes in the plant.
This will allow us to correct for changes in the coupled cavity pole $f_{CC}$, the anti-spring frequency $f_s$ and quality factor $Q$, once these are successfully tracked.
Third, the frequency-dependent systematic errors found from the Gaussian process regressions will be applied directly to the calibrated GW strain data $h(t)$ as it is produced, again through time-domain filters.
The above work would completely eliminate all known systematic errors from our calibrated data.

As we reduce the calibration uncertainty, properly characterizing systematic errors becomes much more important for precision astrophysics.  
Any systematic errors left unaccounted for in the calibrated data can result in systematic errors in binary black hole source parameters, compact binary merger rates, or tests of general relativity.
Our direct measurements of our detector control loop plants combined with the physics-motivated response function model provide a sanity check that our understanding of the interferometer is close to correct.

There are a few considerations requiring quantification at the new low levels of uncertainty.
One is understanding the difference between the quadruple pendulum response to an actual gravitational wave versus its reponse to the photon calibrator.  
In general, we care about the response of the test mass to external displacement, which causes light to be phase shifted out of the inteferometer's antisymmetric port.
We simulate a gravitational wave by pushing on only a single end test mass with the photon calibrator laser.
However, a real gravitational wave stretches space in the entire detector, in particular, the upper stages of the pendulum and the input test masses.
The effect of this difference on calibrated GW data is on the order of about 1\% at 10 Hz, and increases at lower frequencies.
This now must be considered quantitatively as uncertainties approach this level.

Another consideration is the photon calibration actuation strength $H_{PCAL}(t)$.
Currently, the relative uncertainty in $H_{PCAL}(t)$ is 0.79\% \cite{aLIGOPCALPaper}.
This is the fundamental limit on our uncertainty in the response $R$ and therefore the GW strain data $h$.
The uncertainty in $H_{PCAL}(t)$ is dominated by uncertainty in the laser power and test mass rotation \cite{aLIGOPCALPaper}.
To push this fundamental limit lower, better measurements of the photon calibrator laser power and test mass rotation must be made, or more precise methods of calibration outside of the photon calibrator may need to be considered.

The uncertainty budget does not include error from test mass elastic deformation due to the PCAL laser exciting test mass vibrational modes.
Preliminary evidence suggests that above around 3 kHz, elastic deformation has a significant effect on the calibration accuracy.
Elastic deformation due to the PCAL must be further understood, monitored, and included in the uncertainty budget directly.

\section{Conclusion}
\label{sec:Conclusion}

The uncertainty and systematic error estimates reported in this paper represent a comprehensive characterization of our H1 and L1 detector calibrations for observing run two.
In Advanced LIGO's lowest noise region, from about 20 Hz to 1 kHz, the uncertainty in the calibrated data has been reduced from what was previously reported in \cite{GW150914CalPaper}.
The uncertainty estimates for O2 give more refined results, with uncertainty growing at extreme frequency regions below 20 Hz and above 1 kHz, and reduced uncertainty in the low noise frequency region.

GW170104's detection and parameter estimation are primarily limited by noise, and not by calibration uncertainty.
As Advanced LIGO becomes more and more sensitive, the signal-to-noise ratio of some detections will become quite large (as high as 100 or more), and calibration uncertainty will begin contributing significantly to source parameter estimation uncertainty.
With more observing time comes more detections, enabling new tests of general relativity which will be limited by the precision of our detector data.
Precision astrophysics demands the best understanding of our calibrated data possible.  
The methods described in this paper were developed primarily to enable the best science possible from LIGO's gravitational wave detections.

\section{Acknowledgements}
The authors gratefully acknowledge the support of the United States
National Science Foundation (NSF) for the construction and operation of the
LIGO Laboratory and Advanced LIGO as well as the Science and Technology Facilities Council (STFC) of the
United Kingdom, the Max-Planck-Society (MPS), and the State of
Niedersachsen/Germany for support of the construction of Advanced LIGO 
and construction and operation of the GEO600 detector. 
Additional support for Advanced LIGO was provided by the Australian Research Council.
The authors gratefully acknowledge the Italian Istituto Nazionale di Fisica Nucleare (INFN),  
the French Centre National de la Recherche Scientifique (CNRS) and
the Foundation for Fundamental Research on Matter supported by the Netherlands Organisation for Scientific Research, 
for the construction and operation of the Virgo detector
and the creation and support  of the EGO consortium. 
The authors also gratefully acknowledge research support from these agencies as well as by 
the Council of Scientific and Industrial Research of India, 
Department of Science and Technology, India,
Science \& Engineering Research Board (SERB), India,
Ministry of Human Resource Development, India,
the Spanish Ministerio de Econom\'ia y Competitividad,
the Conselleria d'Economia i Competitivitat and Conselleria d'Educaci\'o, Cultura i Universitats of the Govern de les Illes Balears,
the National Science Centre of Poland,
the European Union,
the Royal Society, 
the Scottish Funding Council, 
the Scottish Universities Physics Alliance, 
the Lyon Institute of Origins (LIO),
the National Research Foundation of Korea,
Industry Canada and the Province of Ontario through the Ministry of Economic Development and Innovation, 
the National Science and Engineering Research Council Canada,
the Brazilian Ministry of Science, Technology, and Innovation,
the Research Corporation, 
Ministry of Science and Technology (MOST), Taiwan and the Kavli Foundation.
The authors gratefully acknowledge the support of the NSF, STFC, MPS, INFN, CNRS and the
State of Niedersachsen/Germany for provision of computational resources.
This article has been assigned the LIGO document number P1600139.

\appendix
\section{Observing Run One Results}
\label{sec:appendixA}
During O1, there were two loud gravitational wave detections, GW150914 and GW151226 \cite{GW150914, GW151226}.
There was also a relatively quiet transient, LVT151012, that was likely a gravitational wave \cite{O1BBHPaper}.
The previous report of the calibration uncertainty on these events did not include the new methods developed for O2 \cite{GW150914CalPaper}.
We have returned to the O1 data to recalculate the calibration uncertainties at the time of the three O1 events. 
The old calibration uncertainty method results are reported in Table III of \cite{O1BBHPaper}.
The results shown are for version 2 (C02) of the calibrated GW strain data, available shortly after the run ended in January 2016.

\begin{figure*}
\includegraphics[width=0.48\textwidth]{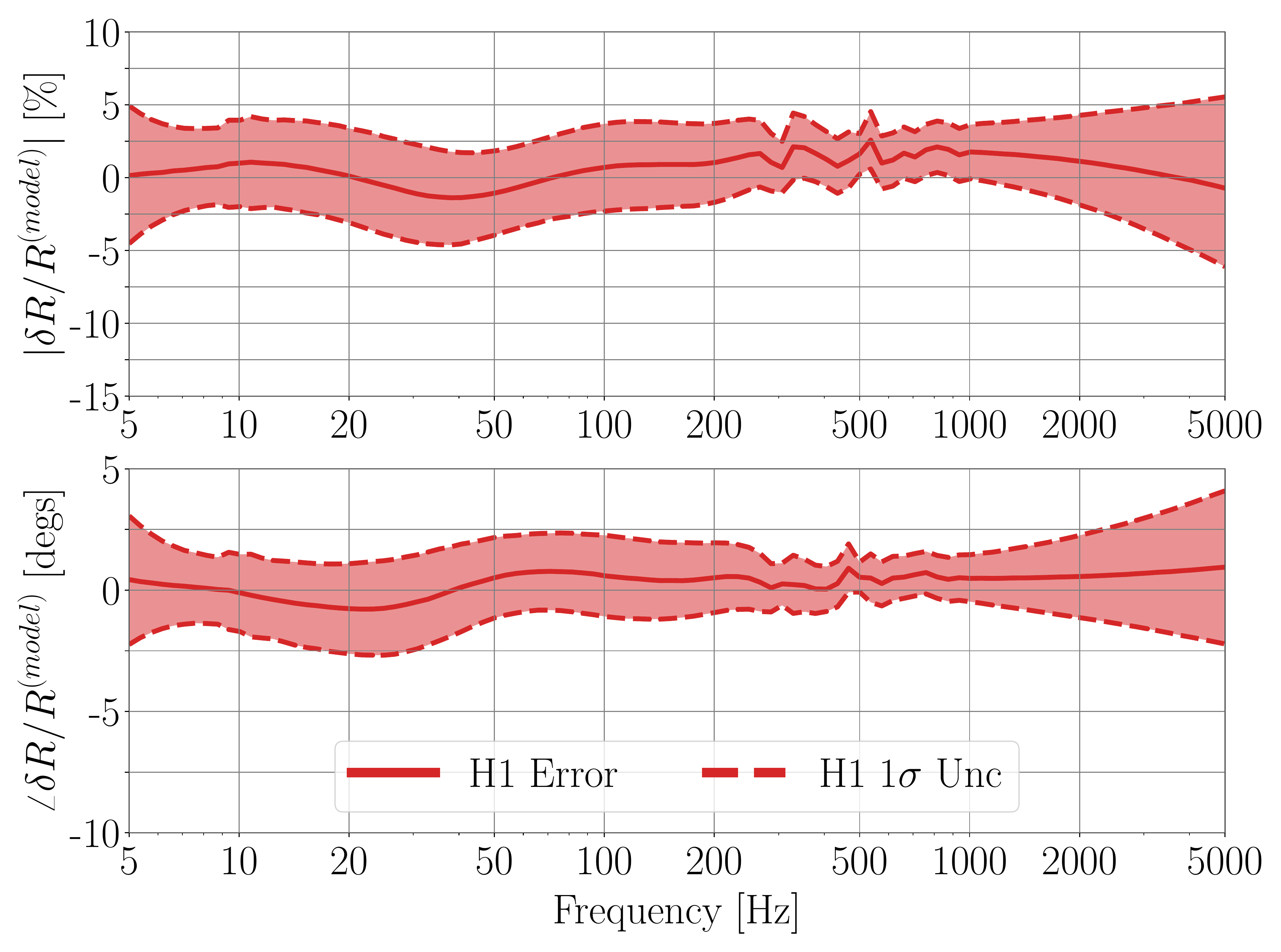}
\includegraphics[width=0.48\textwidth]{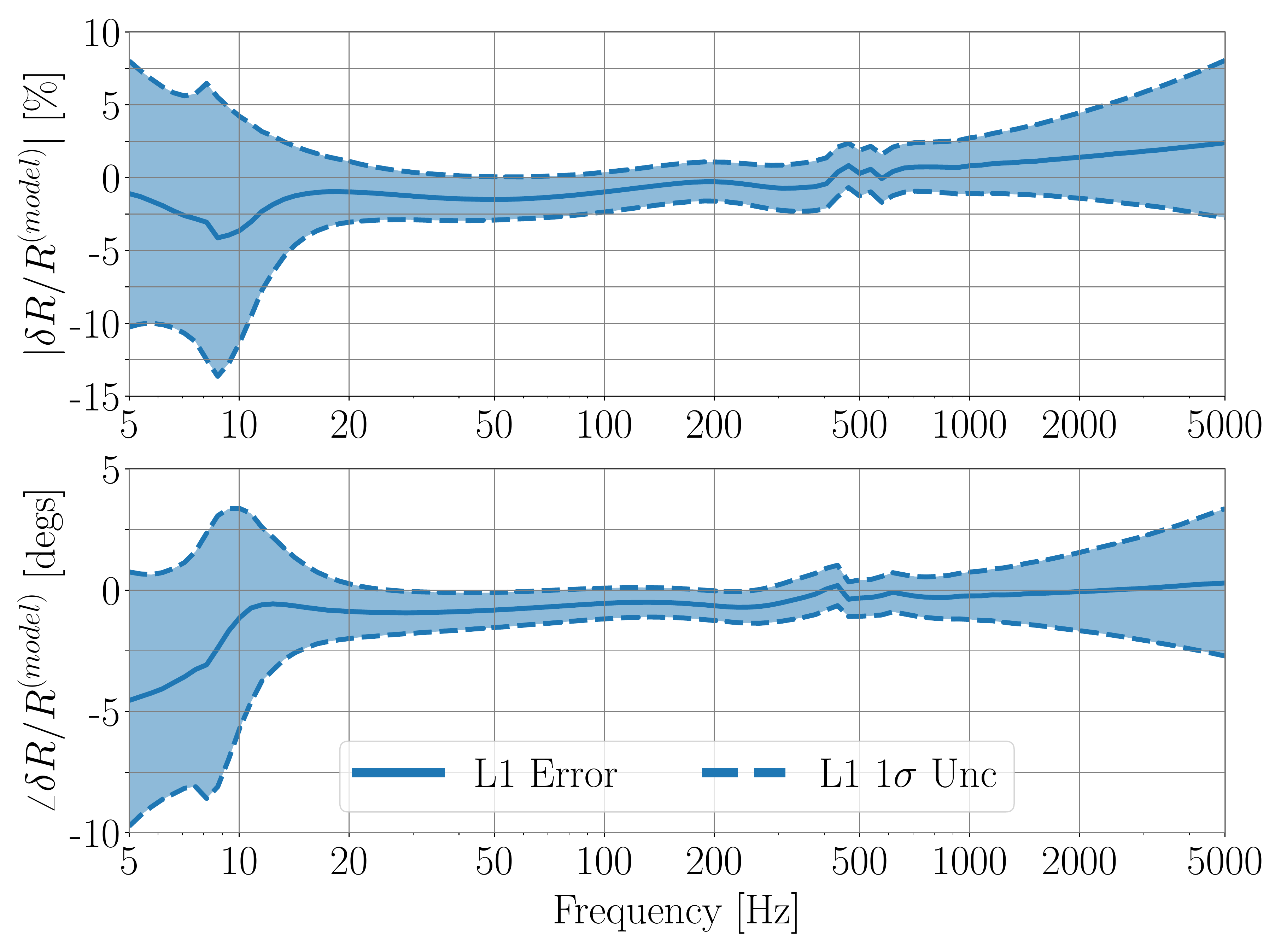}
\includegraphics[width=0.48\textwidth]{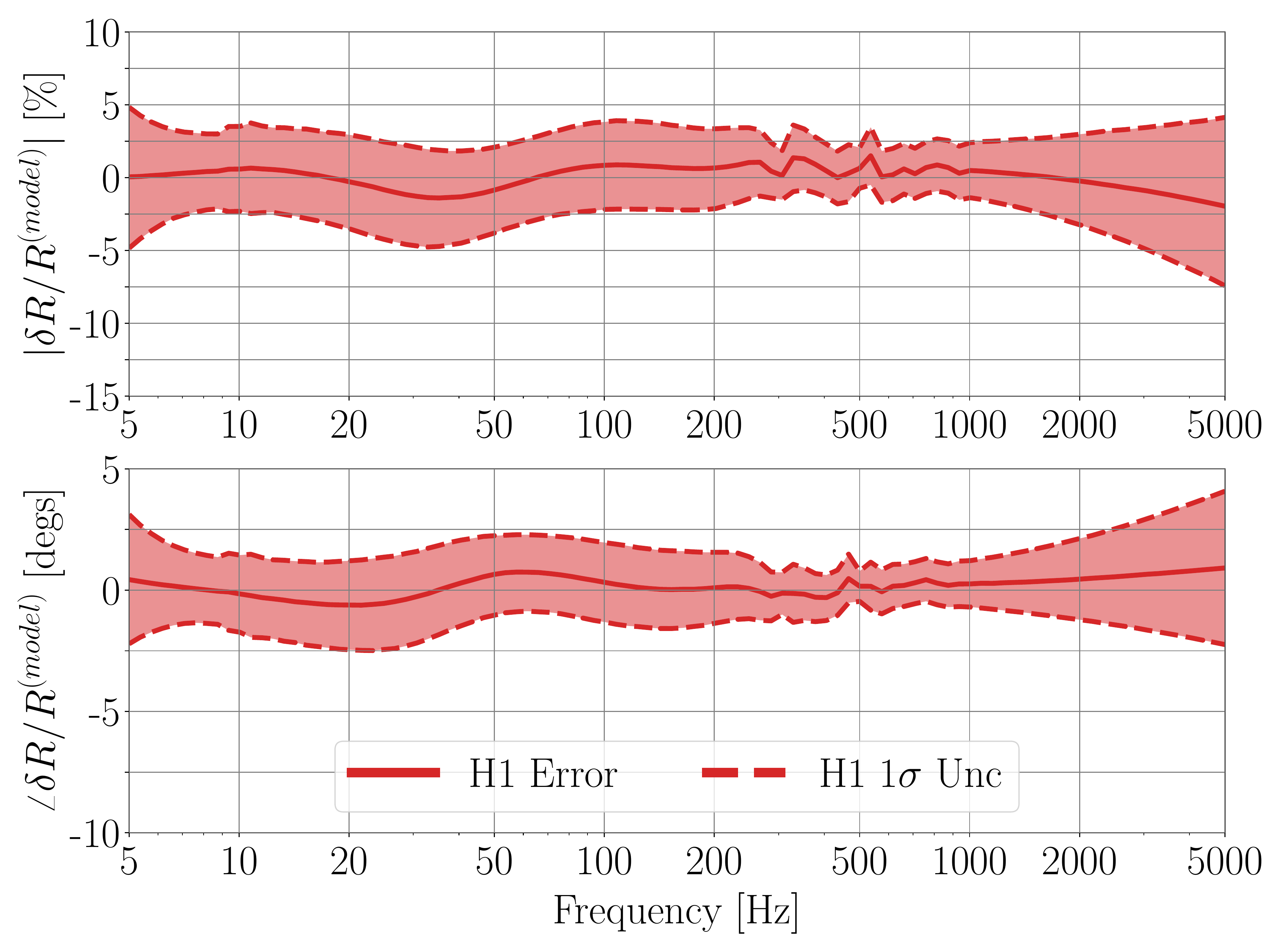}
\includegraphics[width=0.48\textwidth]{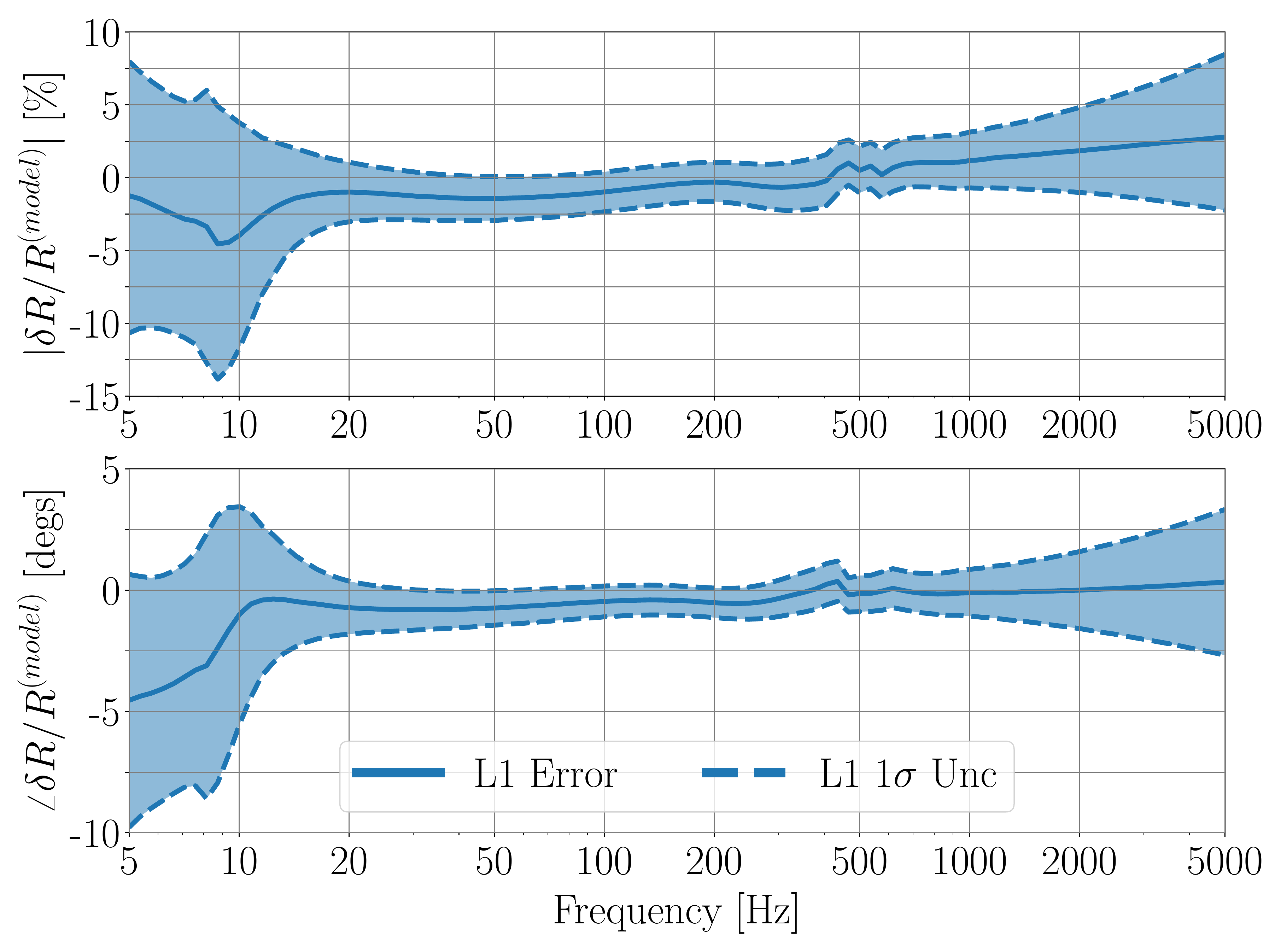}
\includegraphics[width=0.48\textwidth]{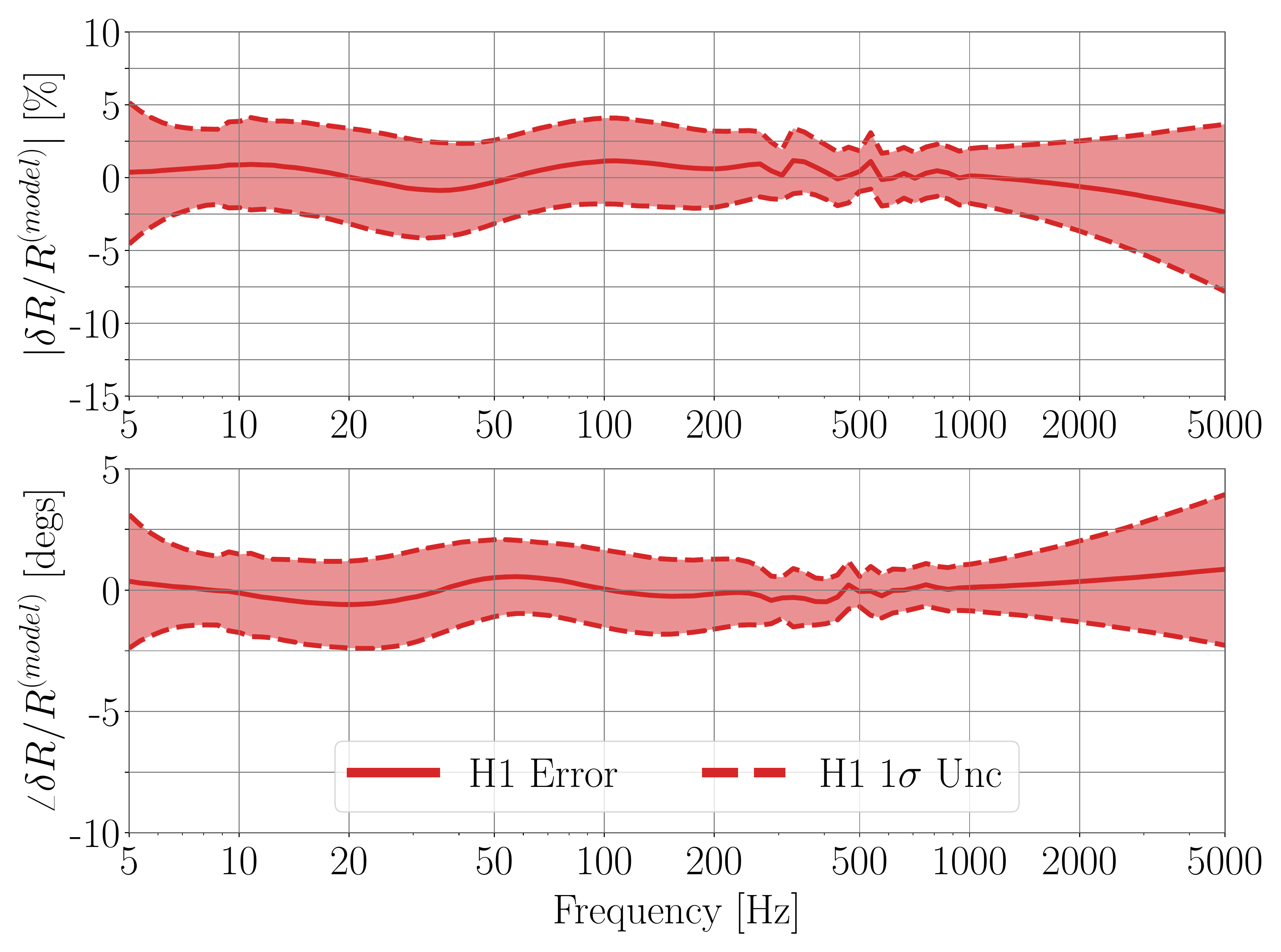}
\includegraphics[width=0.48\textwidth]{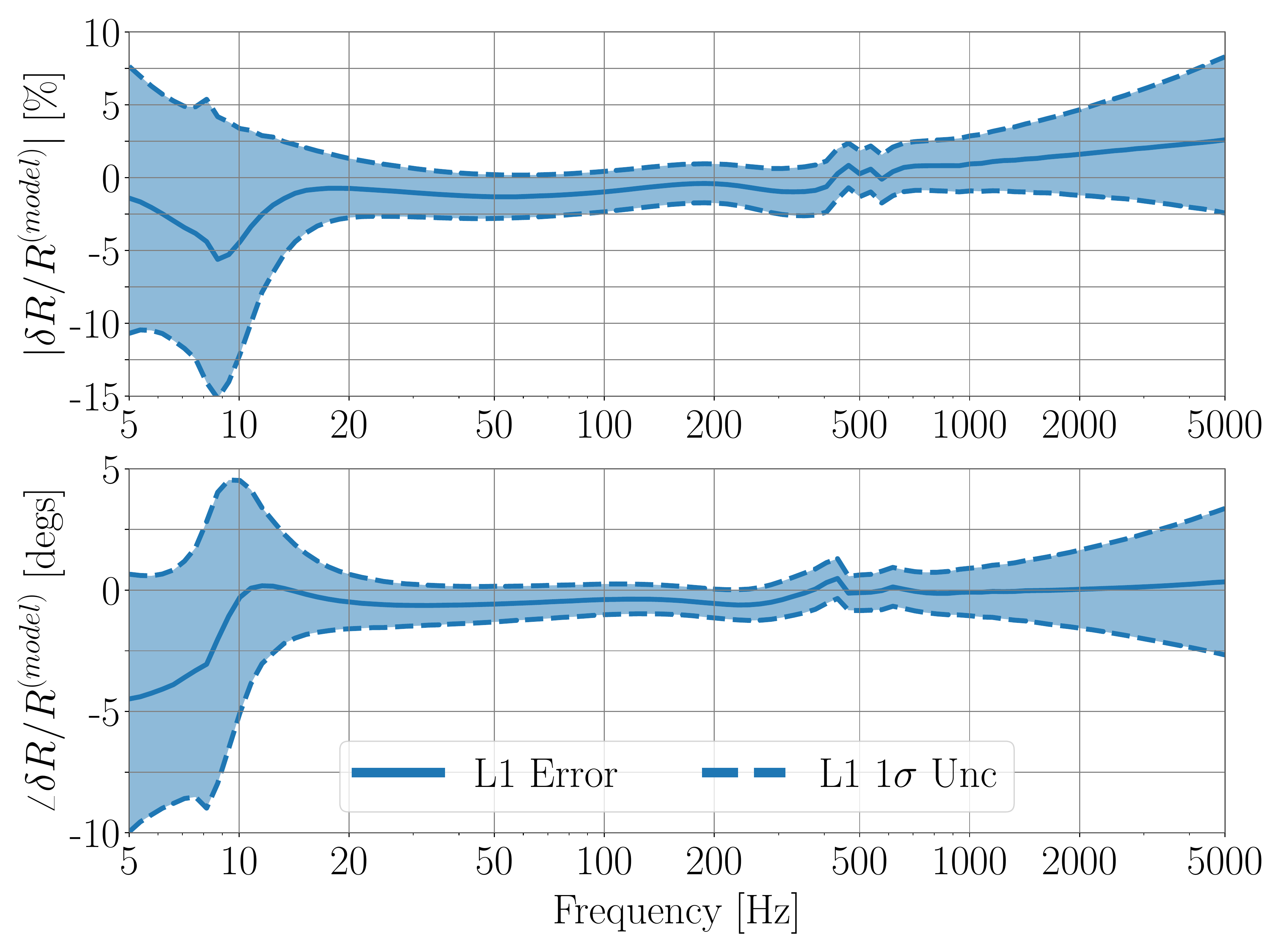}
\caption{Total Calibration Uncertainty Budgets for GW150914, LVT151012, and GW151226. \\
Uncertainties for GW150914 are on top, LVT151012 in the middle, and GW151226 on the bottom.
The uncertainty in the calibrated response function for the H1 detector is on the left, and for L1 is on the right.
The $y$ axis is relative response error $\delta R / R^{(\text{model})}$ and uncertainty $\sigma_{R} / R^{(\text{model})}$, with magnitude on top and phase on the bottom.  
The uncertainties for these O1 events have been calculated using the refined O2 methods described in this paper.
All the budgets are quite similar with slight differences coming from the uncorrected time dependent cavity pole.
The jagged lines around 300-500 Hz come from the actuation function notches, like those seen in Figure \ref{fig:ActuationPlant}.
These budgets report a smaller uncertainty in the region around 100 Hz than reported in past calibration uncertainty publications, with significantly smaller systematic error fluctuations \cite{GW150914CalPaper}.
The refined calibration uncertainty budget methods give a more sensible uncertainty budget for O1 events, with uncertainty expanding at extreme frequencies and reduced in the lowest noise regions.
}
\label{fig:O1ResultsPlot}
\end{figure*}

\begin{figure*}
\includegraphics[width=0.48\textwidth]{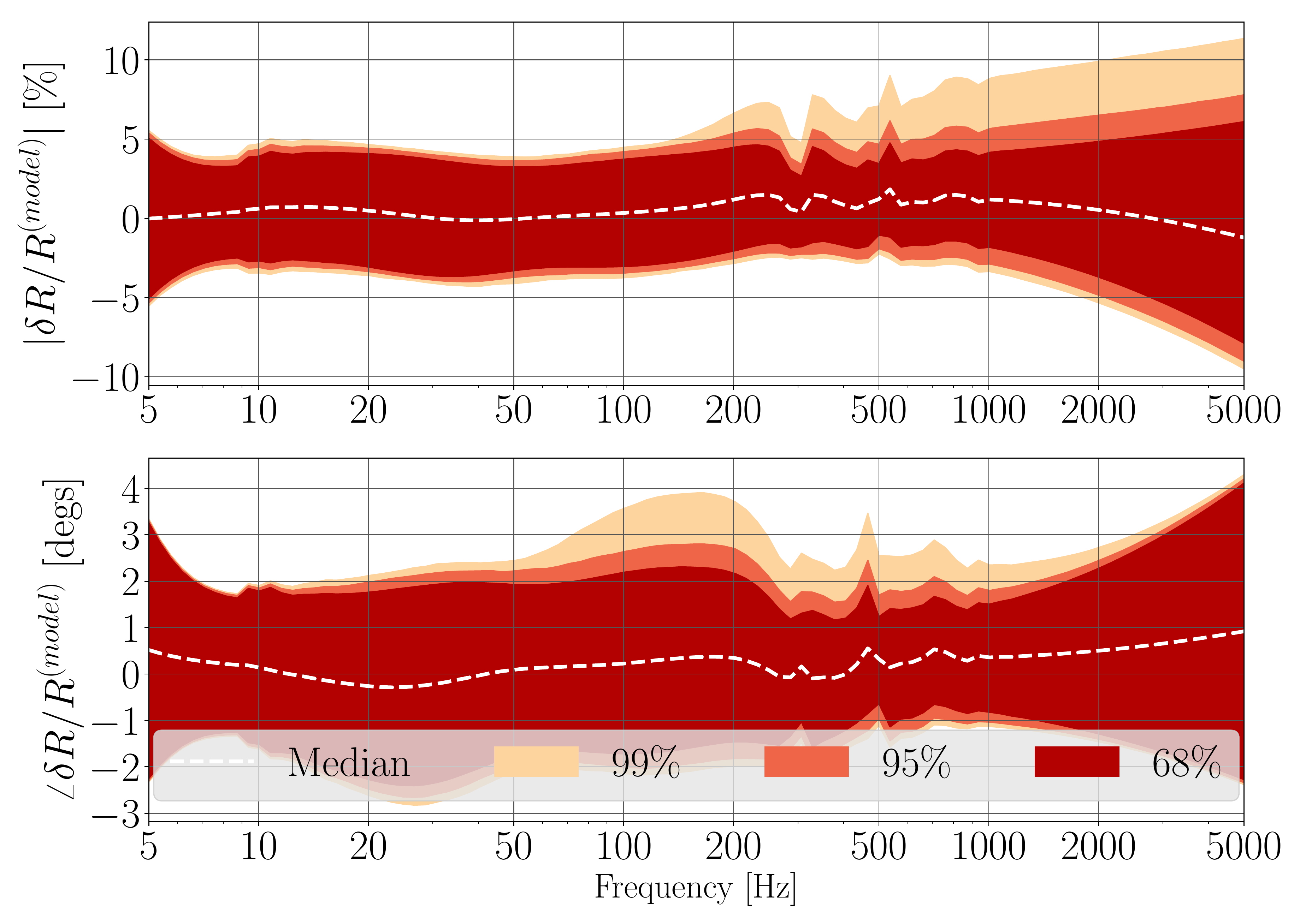}
\includegraphics[width=0.48\textwidth]{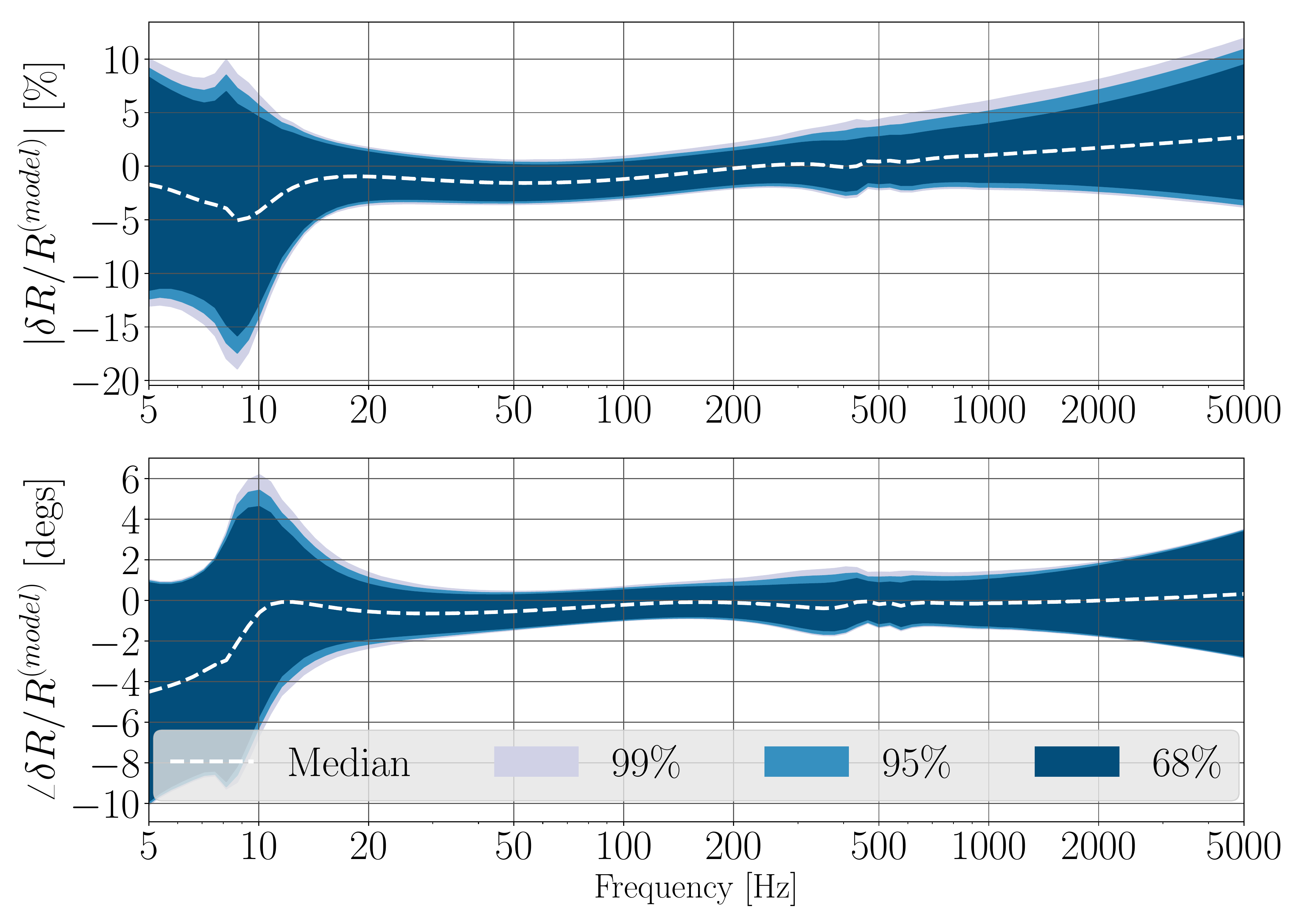}
\caption{Total Calibration Uncertainty Percentiles for All of Observing Run One.\\
The uncertainty in the calibrated response function for the H1 detector is on the left, and for L1 is on the right.
The $y$ axis is relative response error $\delta R / R^{(\text{model})}$ and uncertainty $\sigma_{R} / R^{(\text{model})}$, with magnitude on top and phase on the bottom.  
The dashed white line is the median relative response, while the colors represent the $1\sigma$ calibration uncertainty for 68\%, 95\%, and 99\% of the run's time.
Both interferometer's uncertainty and systematic error are fairly consistent over the course of the run, with the larger fluctuations coming from changes in the coupled cavity pole, which is not corrected for in O1 calibrated data.
}
\label{fig:O1PercentilePlots}
\end{figure*}

Figure \ref{fig:O1ResultsPlot} shows plots of the calibration uncertainties at the times of the O1 events.  Table \ref{tab:O1EventsExtremeUncertainty} reports the extreme uncertainties at the times of the O1 events.
Figure \ref{fig:O1PercentilePlots} shows the calibration uncertainty for all of O1, meaning the GW strain data from September 14, 2015 through January 19, 2016.

\begin{table*}

\caption{GW150914, LVT151012, and GW151226 Extreme Uncertainty \\
Below are the extreme calibration uncertainty values for H1 and L1 at the time of GW150914, LVT151012, and GW151226 in the 20-1024 Hz frequency range.
``Extreme uncertainty'' refers to the maximum and mininum of error $\pm 1\sigma$ uncertainty.
The plots informing these tables can be seen in Figure \ref{fig:O1ResultsPlot}
} \label{tab:O1EventsExtremeUncertainty}
\begin{minipage}[t]{0.3\linewidth} 
\def\arraystretch{1.5}
\begin{tabular}[t]{ c c c }
GW150914 Uncertainty & H1 & L1 \\ 
\hline
$+1\sigma$ Magnitude [\%]    & 3.7 \%    &  1.1 \% \\
$-1\sigma$ Magnitude [\%]     & -4.6 \%  & -3.0 \% \\
$+1\sigma$ Phase [degrees]  & 2.4$^{\circ}$     &  0.3$^{\circ}$  \\
$-1\sigma$ Phase [degrees]   & -2.7$^{\circ}$    & -2.0$^{\circ}$ \\
\hline
\end{tabular}
\end{minipage}
\begin{minipage}[t]{0.3\linewidth}
\def\arraystretch{1.5}
\begin{tabular}[t]{ c c c }
LVT151012 Uncertainty & H1 & L1 \\ 
\hline
$+1\sigma$ Magnitude [\%]    & 3.8 \%    &  1.0 \% \\
$-1\sigma$ Magnitude [\%]     & -4.8 \%  & -3.0 \% \\
$+1\sigma$ Phase [degrees]  & 2.3$^{\circ}$     &  0.4$^{\circ}$  \\
$-1\sigma$ Phase [degrees]   & -2.5$^{\circ}$    & -1.8$^{\circ}$ \\
\hline
\end{tabular}
\end{minipage}
\begin{minipage}[t]{0.3\linewidth}
\def\arraystretch{1.5}
\begin{tabular}[t]{ c c c }
GW151226 Uncertainty & H1 & L1 \\ 
\hline
$+1\sigma$ Magnitude [\%]    & 4.1 \%    &  1.3 \% \\
$-1\sigma$ Magnitude [\%]     & -4.1 \%  & -2.8 \% \\
$+1\sigma$ Phase [degrees]  & 2.1$^{\circ}$     &  0.6$^{\circ}$  \\
$-1\sigma$ Phase [degrees]   & -2.4$^{\circ}$    & -1.6$^{\circ}$ \\
\hline
\end{tabular}
\end{minipage}
\end{table*}

\bibliographystyle{apsrev4-1}

\begin{thebibliography}{35}%
\makeatletter
\providecommand \@ifxundefined [1]{%
 \@ifx{#1\undefined}
}%
\providecommand \@ifnum [1]{%
 \ifnum #1\expandafter \@firstoftwo
 \else \expandafter \@secondoftwo
 \fi
}%
\providecommand \@ifx [1]{%
 \ifx #1\expandafter \@firstoftwo
 \else \expandafter \@secondoftwo
 \fi
}%
\providecommand \natexlab [1]{#1}%
\providecommand \enquote  [1]{``#1''}%
\providecommand \bibnamefont  [1]{#1}%
\providecommand \bibfnamefont [1]{#1}%
\providecommand \citenamefont [1]{#1}%
\providecommand \href@noop [0]{\@secondoftwo}%
\providecommand \href [0]{\begingroup \@sanitize@url \@href}%
\providecommand \@href[1]{\@@startlink{#1}\@@href}%
\providecommand \@@href[1]{\endgroup#1\@@endlink}%
\providecommand \@sanitize@url [0]{\catcode `\\12\catcode `\$12\catcode
  `\&12\catcode `\#12\catcode `\^12\catcode `\_12\catcode `\%12\relax}%
\providecommand \@@startlink[1]{}%
\providecommand \@@endlink[0]{}%
\providecommand \url  [0]{\begingroup\@sanitize@url \@url }%
\providecommand \@url [1]{\endgroup\@href {#1}{\urlprefix }}%
\providecommand \urlprefix  [0]{URL }%
\providecommand \Eprint [0]{\href }%
\providecommand \doibase [0]{http://dx.doi.org/}%
\providecommand \selectlanguage [0]{\@gobble}%
\providecommand \bibinfo  [0]{\@secondoftwo}%
\providecommand \bibfield  [0]{\@secondoftwo}%
\providecommand \translation [1]{[#1]}%
\providecommand \BibitemOpen [0]{}%
\providecommand \bibitemStop [0]{}%
\providecommand \bibitemNoStop [0]{.\EOS\space}%
\providecommand \EOS [0]{\spacefactor3000\relax}%
\providecommand \BibitemShut  [1]{\csname bibitem#1\endcsname}%
\let\auto@bib@innerbib\@empty
\bibitem [{\citenamefont {Abbott}\ \emph
  {et~al.}(2016{\natexlab{a}})\citenamefont {Abbott} \emph
  {et~al.}}]{GW150914}%
  \BibitemOpen
  \bibfield  {author} {\bibinfo {author} {\bibfnamefont {B.~P.}\ \bibnamefont
  {Abbott}} \emph {et~al.} (\bibinfo {collaboration} {LIGO Scientific
  Collaboration and Virgo Collaboration}),\ }\href {\doibase
  10.1103/PhysRevLett.116.061102} {\bibfield  {journal} {\bibinfo  {journal}
  {Phys. Rev. Lett.}\ }\textbf {\bibinfo {volume} {116}},\ \bibinfo {pages}
  {061102} (\bibinfo {year} {2016}{\natexlab{a}})}\BibitemShut {NoStop}%
\bibitem [{\citenamefont {Abbott}\ \emph
  {et~al.}(2016{\natexlab{b}})\citenamefont {Abbott} \emph
  {et~al.}}]{GW151226}%
  \BibitemOpen
  \bibfield  {author} {\bibinfo {author} {\bibfnamefont {B.~P.}\ \bibnamefont
  {Abbott}} \emph {et~al.} (\bibinfo {collaboration} {LIGO Scientific
  Collaboration and Virgo Collaboration}),\ }\href {\doibase
  10.1103/PhysRevLett.116.241103} {\bibfield  {journal} {\bibinfo  {journal}
  {Phys. Rev. Lett.}\ }\textbf {\bibinfo {volume} {116}},\ \bibinfo {pages}
  {241103} (\bibinfo {year} {2016}{\natexlab{b}})}\BibitemShut {NoStop}%
\bibitem [{\citenamefont {Abbott}\ \emph
  {et~al.}(2017{\natexlab{a}})\citenamefont {Abbott} \emph
  {et~al.}}]{GW170104}%
  \BibitemOpen
  \bibfield  {author} {\bibinfo {author} {\bibfnamefont {B.~P.}\ \bibnamefont
  {Abbott}} \emph {et~al.} (\bibinfo {collaboration} {LIGO Scientific and Virgo
  Collaboration}),\ }\href {\doibase 10.1103/PhysRevLett.118.221101} {\bibfield
   {journal} {\bibinfo  {journal} {Phys. Rev. Lett.}\ }\textbf {\bibinfo
  {volume} {118}},\ \bibinfo {pages} {221101} (\bibinfo {year}
  {2017}{\natexlab{a}})}\BibitemShut {NoStop}%
\bibitem [{\citenamefont {Abbott}\ \emph
  {et~al.}(2016{\natexlab{c}})\citenamefont {Abbott} \emph
  {et~al.}}]{O1BBHPaper}%
  \BibitemOpen
  \bibfield  {author} {\bibinfo {author} {\bibfnamefont {B.~P.}\ \bibnamefont
  {Abbott}} \emph {et~al.} (\bibinfo {collaboration} {LIGO Scientific
  Collaboration and Virgo Collaboration}),\ }\href {\doibase
  10.1103/PhysRevX.6.041015} {\bibfield  {journal} {\bibinfo  {journal} {Phys.
  Rev. X}\ }\textbf {\bibinfo {volume} {6}},\ \bibinfo {pages} {041015}
  (\bibinfo {year} {2016}{\natexlab{c}})}\BibitemShut {NoStop}%
\bibitem [{\citenamefont {Abbott}\ \emph
  {et~al.}(2016{\natexlab{d}})\citenamefont {Abbott} \emph
  {et~al.}}]{GW150914PEPaper}%
  \BibitemOpen
  \bibfield  {author} {\bibinfo {author} {\bibfnamefont {B.~P.}\ \bibnamefont
  {Abbott}} \emph {et~al.} (\bibinfo {collaboration} {LIGO Scientific
  Collaboration and Virgo Collaboration}),\ }\href {\doibase
  10.1103/PhysRevLett.116.241102} {\bibfield  {journal} {\bibinfo  {journal}
  {Phys. Rev. Lett.}\ }\textbf {\bibinfo {volume} {116}},\ \bibinfo {pages}
  {241102} (\bibinfo {year} {2016}{\natexlab{d}})}\BibitemShut {NoStop}%
\bibitem [{\citenamefont {Abbott}\ \emph
  {et~al.}(2016{\natexlab{e}})\citenamefont {Abbott} \emph
  {et~al.}}]{RatesPaper}%
  \BibitemOpen
  \bibfield  {author} {\bibinfo {author} {\bibfnamefont {B.~P.}\ \bibnamefont
  {Abbott}} \emph {et~al.},\ }\href
  {http://iopscience.iop.org/article/10.3847/2041-8205/833/1/L1/} {\bibfield
  {journal} {\bibinfo  {journal} {Astrophys. J. Lett.}\ }\textbf {\bibinfo
  {volume} {833}},\ \bibinfo {pages} {L1} (\bibinfo {year}
  {2016}{\natexlab{e}})}\BibitemShut {NoStop}%
\bibitem [{\citenamefont {Abbott}\ \emph
  {et~al.}(2016{\natexlab{f}})\citenamefont {Abbott} \emph
  {et~al.}}]{GW150915GRTests}%
  \BibitemOpen
  \bibfield  {author} {\bibinfo {author} {\bibfnamefont {B.~P.}\ \bibnamefont
  {Abbott}} \emph {et~al.} (\bibinfo {collaboration} {LIGO Scientific and Virgo
  Collaborations}),\ }\href {\doibase 10.1103/PhysRevLett.116.221101}
  {\bibfield  {journal} {\bibinfo  {journal} {Phys. Rev. Lett.}\ }\textbf
  {\bibinfo {volume} {116}},\ \bibinfo {pages} {221101} (\bibinfo {year}
  {2016}{\natexlab{f}})}\BibitemShut {NoStop}%
\bibitem [{\citenamefont {Yunes}\ \emph {et~al.}(2016)\citenamefont {Yunes},
  \citenamefont {Yagi},\ and\ \citenamefont {Pretorius}}]{TestingGR2016}%
  \BibitemOpen
  \bibfield  {author} {\bibinfo {author} {\bibfnamefont {N.}~\bibnamefont
  {Yunes}}, \bibinfo {author} {\bibfnamefont {K.}~\bibnamefont {Yagi}}, \ and\
  \bibinfo {author} {\bibfnamefont {F.}~\bibnamefont {Pretorius}},\ }\href
  {\doibase 10.1103/PhysRevD.94.084002} {\bibfield  {journal} {\bibinfo
  {journal} {Phys. Rev. D}\ }\textbf {\bibinfo {volume} {94}},\ \bibinfo
  {pages} {084002} (\bibinfo {year} {2016})}\BibitemShut {NoStop}%
\bibitem [{\citenamefont {Chamberlain}\ and\ \citenamefont
  {Yunes}(2017)}]{TestingGR2017}%
  \BibitemOpen
  \bibfield  {author} {\bibinfo {author} {\bibfnamefont {K.}~\bibnamefont
  {Chamberlain}}\ and\ \bibinfo {author} {\bibfnamefont {N.}~\bibnamefont
  {Yunes}},\ }\href {https://arxiv.org/abs/1704.08268} {\bibfield  {journal}
  {\bibinfo  {journal} {arXiv preprint arXiv:1704.08268}\ } (\bibinfo {year}
  {2017})}\BibitemShut {NoStop}%
\bibitem [{\citenamefont {Abbott}\ \emph
  {et~al.}(2017{\natexlab{b}})\citenamefont {Abbott} \emph
  {et~al.}}]{O1CW2017a}%
  \BibitemOpen
  \bibfield  {author} {\bibinfo {author} {\bibfnamefont {B.~P.}\ \bibnamefont
  {Abbott}} \emph {et~al.},\ }\href
  {http://iopscience.iop.org/article/10.3847/1538-4357/aa677f} {\bibfield
  {journal} {\bibinfo  {journal} {The Astrophysical Journal}\ }\textbf
  {\bibinfo {volume} {839}},\ \bibinfo {pages} {12} (\bibinfo {year}
  {2017}{\natexlab{b}})}\BibitemShut {NoStop}%
\bibitem [{\citenamefont {Abbott}\ \emph
  {et~al.}(2017{\natexlab{c}})\citenamefont {Abbott} \emph
  {et~al.}}]{O1CW2017b}%
  \BibitemOpen
  \bibfield  {author} {\bibinfo {author} {\bibfnamefont {B.~P.}\ \bibnamefont
  {Abbott}} \emph {et~al.},\ }\href {https://dcc.ligo.org/LIGO-P1700127}
  {\bibfield  {journal} {\bibinfo  {journal} {in prep}\ } (\bibinfo {year}
  {2017}{\natexlab{c}})}\BibitemShut {NoStop}%
\bibitem [{\citenamefont {Del~Pozzo}\ \emph {et~al.}(2017)\citenamefont
  {Del~Pozzo}, \citenamefont {Li},\ and\ \citenamefont
  {Messenger}}]{Cosmology}%
  \BibitemOpen
  \bibfield  {author} {\bibinfo {author} {\bibfnamefont {W.}~\bibnamefont
  {Del~Pozzo}}, \bibinfo {author} {\bibfnamefont {T.~G.~F.}\ \bibnamefont
  {Li}}, \ and\ \bibinfo {author} {\bibfnamefont {C.}~\bibnamefont
  {Messenger}},\ }\href {\doibase 10.1103/PhysRevD.95.043502} {\bibfield
  {journal} {\bibinfo  {journal} {Phys. Rev. D}\ }\textbf {\bibinfo {volume}
  {95}},\ \bibinfo {pages} {043502} (\bibinfo {year} {2017})}\BibitemShut
  {NoStop}%
\bibitem [{\citenamefont {Abbott}\ \emph
  {et~al.}(2017{\natexlab{d}})\citenamefont {Abbott} \emph
  {et~al.}}]{O1Stoch2017a}%
  \BibitemOpen
  \bibfield  {author} {\bibinfo {author} {\bibfnamefont {B.~P.}\ \bibnamefont
  {Abbott}} \emph {et~al.} (\bibinfo {collaboration} {LIGO Scientific
  Collaboration and Virgo Collaboration}),\ }\href {\doibase
  10.1103/PhysRevLett.118.121101} {\bibfield  {journal} {\bibinfo  {journal}
  {Phys. Rev. Lett.}\ }\textbf {\bibinfo {volume} {118}},\ \bibinfo {pages}
  {121101} (\bibinfo {year} {2017}{\natexlab{d}})}\BibitemShut {NoStop}%
\bibitem [{\citenamefont {Abbott}\ \emph
  {et~al.}(2017{\natexlab{e}})\citenamefont {Abbott} \emph
  {et~al.}}]{O1Stoch2017b}%
  \BibitemOpen
  \bibfield  {author} {\bibinfo {author} {\bibfnamefont {B.~P.}\ \bibnamefont
  {Abbott}} \emph {et~al.} (\bibinfo {collaboration} {LIGO Scientific
  Collaboration and Virgo Collaboration}),\ }\href {\doibase
  10.1103/PhysRevLett.118.121102} {\bibfield  {journal} {\bibinfo  {journal}
  {Phys. Rev. Lett.}\ }\textbf {\bibinfo {volume} {118}},\ \bibinfo {pages}
  {121102} (\bibinfo {year} {2017}{\natexlab{e}})}\BibitemShut {NoStop}%
\bibitem [{\citenamefont {Chernoff}\ and\ \citenamefont
  {Finn}(1993)}]{Chernoff1993}%
  \BibitemOpen
  \bibfield  {author} {\bibinfo {author} {\bibfnamefont {D.~F.}\ \bibnamefont
  {Chernoff}}\ and\ \bibinfo {author} {\bibfnamefont {L.~S.}\ \bibnamefont
  {Finn}},\ }\href {http://adsabs.harvard.edu/doi/10.1086/186898} {\bibfield
  {journal} {\bibinfo  {journal} {Astrophys. J.}\ }\textbf {\bibinfo {volume}
  {411}},\ \bibinfo {pages} {L5} (\bibinfo {year} {1993})}\BibitemShut
  {NoStop}%
\bibitem [{\citenamefont {Abbott}\ \emph
  {et~al.}(2017{\natexlab{f}})\citenamefont {Abbott} \emph
  {et~al.}}]{GW150914CalPaper}%
  \BibitemOpen
  \bibfield  {author} {\bibinfo {author} {\bibfnamefont {B.~P.}\ \bibnamefont
  {Abbott}} \emph {et~al.} (\bibinfo {collaboration} {LIGO Scientific
  Collaboration}),\ }\href {\doibase 10.1103/PhysRevD.95.062003} {\bibfield
  {journal} {\bibinfo  {journal} {Phys. Rev. D}\ }\textbf {\bibinfo {volume}
  {95}},\ \bibinfo {pages} {062003} (\bibinfo {year}
  {2017}{\natexlab{f}})}\BibitemShut {NoStop}%
\bibitem [{\citenamefont {Aasi}\ \emph {et~al.}(2015)\citenamefont {Aasi} \emph
  {et~al.}}]{AdvLIGOPaper}%
  \BibitemOpen
  \bibfield  {author} {\bibinfo {author} {\bibfnamefont {J.}~\bibnamefont
  {Aasi}} \emph {et~al.},\ }\href
  {http://iopscience.iop.org/article/10.1088/0264-9381/32/7/074001} {\bibfield
  {journal} {\bibinfo  {journal} {Classical Quant. Grav.}\ }\textbf {\bibinfo
  {volume} {32}},\ \bibinfo {pages} {074001} (\bibinfo {year}
  {2015})}\BibitemShut {NoStop}%
\bibitem [{\citenamefont {Abbott}\ \emph
  {et~al.}(2016{\natexlab{g}})\citenamefont {Abbott} \emph
  {et~al.}}]{GW150914DetectorPaper}%
  \BibitemOpen
  \bibfield  {author} {\bibinfo {author} {\bibfnamefont {B.~P.}\ \bibnamefont
  {Abbott}} \emph {et~al.} (\bibinfo {collaboration} {LIGO Scientific
  Collaboration and Virgo Collaboration}),\ }\href {\doibase
  10.1103/PhysRevLett.116.131103} {\bibfield  {journal} {\bibinfo  {journal}
  {Phys. Rev. Lett.}\ }\textbf {\bibinfo {volume} {116}},\ \bibinfo {pages}
  {131103} (\bibinfo {year} {2016}{\natexlab{g}})}\BibitemShut {NoStop}%
\bibitem [{\citenamefont {Izumi}\ and\ \citenamefont
  {Sigg}(2016)}]{aLIGOLSCPaper}%
  \BibitemOpen
  \bibfield  {author} {\bibinfo {author} {\bibfnamefont {K.}~\bibnamefont
  {Izumi}}\ and\ \bibinfo {author} {\bibfnamefont {D.}~\bibnamefont {Sigg}},\
  }\href {http://iopscience.iop.org/article/10.1088/0264-9381/34/1/015001}
  {\bibfield  {journal} {\bibinfo  {journal} {Classical Quant. Grav.}\ }\textbf
  {\bibinfo {volume} {34}},\ \bibinfo {pages} {015001} (\bibinfo {year}
  {2016})}\BibitemShut {NoStop}%
\bibitem [{\citenamefont {Robertson}\ \emph {et~al.}(2002)\citenamefont
  {Robertson} \emph {et~al.}}]{Suspensions2002}%
  \BibitemOpen
  \bibfield  {author} {\bibinfo {author} {\bibfnamefont {N.~A.}\ \bibnamefont
  {Robertson}} \emph {et~al.},\ }\href
  {http://iopscience.iop.org/article/10.1088/0264-9381/19/15/311} {\bibfield
  {journal} {\bibinfo  {journal} {Classical Quant. Grav.}\ }\textbf {\bibinfo
  {volume} {19}},\ \bibinfo {pages} {4043} (\bibinfo {year}
  {2002})}\BibitemShut {NoStop}%
\bibitem [{\citenamefont {Aston}\ \emph {et~al.}(2012)\citenamefont {Aston}
  \emph {et~al.}}]{Suspensions2012}%
  \BibitemOpen
  \bibfield  {author} {\bibinfo {author} {\bibfnamefont {S.~M.}\ \bibnamefont
  {Aston}} \emph {et~al.},\ }\href
  {http://iopscience.iop.org/article/10.1088/0264-9381/29/23/235004} {\bibfield
   {journal} {\bibinfo  {journal} {Classical Quant. Grav.}\ }\textbf {\bibinfo
  {volume} {29}},\ \bibinfo {pages} {235004} (\bibinfo {year}
  {2012})}\BibitemShut {NoStop}%
\bibitem [{\citenamefont {Matichard}\ \emph {et~al.}(2015)\citenamefont
  {Matichard} \emph {et~al.}}]{aLIGOSEI}%
  \BibitemOpen
  \bibfield  {author} {\bibinfo {author} {\bibfnamefont {F.}~\bibnamefont
  {Matichard}} \emph {et~al.},\ }\href
  {http://www.sciencedirect.com/science/article/pii/S0141635914001561}
  {\bibfield  {journal} {\bibinfo  {journal} {Precision Engineering}\ }\textbf
  {\bibinfo {volume} {40}},\ \bibinfo {pages} {273} (\bibinfo {year}
  {2015})}\BibitemShut {NoStop}%
\bibitem [{\citenamefont {Carbone}\ \emph {et~al.}(2012)\citenamefont {Carbone}
  \emph {et~al.}}]{QUADSensorsAndActuators}%
  \BibitemOpen
  \bibfield  {author} {\bibinfo {author} {\bibfnamefont {L.}~\bibnamefont
  {Carbone}} \emph {et~al.},\ }\href
  {http://iopscience.iop.org/article/10.1088/0264-9381/29/11/115005} {\bibfield
   {journal} {\bibinfo  {journal} {Classical Quant. Grav.}\ }\textbf {\bibinfo
  {volume} {29}},\ \bibinfo {pages} {115005} (\bibinfo {year}
  {2012})}\BibitemShut {NoStop}%
\bibitem [{\citenamefont {Bendat}\ and\ \citenamefont
  {Piersol}(2011)}]{BendatPiersolCoherenceUncertainty}%
  \BibitemOpen
  \bibfield  {author} {\bibinfo {author} {\bibfnamefont {J.~S.}\ \bibnamefont
  {Bendat}}\ and\ \bibinfo {author} {\bibfnamefont {A.~G.}\ \bibnamefont
  {Piersol}},\ }\href@noop {} {\emph {\bibinfo {title} {Random data: analysis
  and measurement procedures}}},\ Vol.\ \bibinfo {volume} {729}\ (\bibinfo
  {publisher} {John Wiley \& Sons},\ \bibinfo {year} {2011})\BibitemShut
  {NoStop}%
\bibitem [{\citenamefont {Buonanno}\ and\ \citenamefont
  {Chen}(2002)}]{BuonnanoChen2002}%
  \BibitemOpen
  \bibfield  {author} {\bibinfo {author} {\bibfnamefont {A.}~\bibnamefont
  {Buonanno}}\ and\ \bibinfo {author} {\bibfnamefont {Y.}~\bibnamefont
  {Chen}},\ }\href {\doibase 10.1103/PhysRevD.65.042001} {\bibfield  {journal}
  {\bibinfo  {journal} {Phys. Rev. D}\ }\textbf {\bibinfo {volume} {65}},\
  \bibinfo {pages} {042001} (\bibinfo {year} {2002})}\BibitemShut {NoStop}%
\bibitem [{\citenamefont {Ward}(2010)}]{WardThesis}%
  \BibitemOpen
  \bibfield  {author} {\bibinfo {author} {\bibfnamefont {R.~L.}\ \bibnamefont
  {Ward}},\ }\emph {\bibinfo {title} {Length sensing and control of a prototype
  advanced interferometric gravitational wave detector}},\ \href
  {http://thesis.library.caltech.edu/5836/2/main_final_withheld.pdf} {Ph.D.
  thesis},\ \bibinfo  {school} {California Institute of Technology} (\bibinfo
  {year} {2010})\BibitemShut {NoStop}%
\bibitem [{\citenamefont {Hall}(2017)}]{HallThesis}%
  \BibitemOpen
  \bibfield  {author} {\bibinfo {author} {\bibfnamefont {E.~D.}\ \bibnamefont
  {Hall}},\ }\emph {\bibinfo {title} {Long-Baseline Laser Interferometry for
  the Detection of Binary Black-Hole Mergers}},\ \href
  {http://thesis.library.caltech.edu/10031/1/evan-hall-thesis.pdf} {Ph.D.
  thesis},\ \bibinfo  {school} {California Institute of Technology} (\bibinfo
  {year} {2017})\BibitemShut {NoStop}%
\bibitem [{\citenamefont {Rakhmanov}\ \emph {et~al.}(2002)\citenamefont
  {Rakhmanov}, \citenamefont {Savage}, \citenamefont {Reitze},\ and\
  \citenamefont {Tanner}}]{Rakhmanov2002}%
  \BibitemOpen
  \bibfield  {author} {\bibinfo {author} {\bibfnamefont {M.}~\bibnamefont
  {Rakhmanov}}, \bibinfo {author} {\bibfnamefont {R.}~\bibnamefont {Savage}},
  \bibinfo {author} {\bibfnamefont {D.}~\bibnamefont {Reitze}}, \ and\ \bibinfo
  {author} {\bibfnamefont {D.}~\bibnamefont {Tanner}},\ }\href
  {http://www.sciencedirect.com/science/article/pii/S037596010201469X}
  {\bibfield  {journal} {\bibinfo  {journal} {Physics Letters A}\ }\textbf
  {\bibinfo {volume} {305}},\ \bibinfo {pages} {239} (\bibinfo {year}
  {2002})}\BibitemShut {NoStop}%
\bibitem [{\citenamefont {Karki}\ \emph {et~al.}(2016)\citenamefont {Karki}
  \emph {et~al.}}]{aLIGOPCALPaper}%
  \BibitemOpen
  \bibfield  {author} {\bibinfo {author} {\bibfnamefont {S.}~\bibnamefont
  {Karki}} \emph {et~al.},\ }\href
  {http://aip.scitation.org/doi/abs/10.1063/1.4967303} {\bibfield  {journal}
  {\bibinfo  {journal} {Rev. Sci. Instrum.}\ }\textbf {\bibinfo {volume}
  {87}},\ \bibinfo {pages} {114503} (\bibinfo {year} {2016})}\BibitemShut
  {NoStop}%
\bibitem [{\citenamefont {Tuyenbayev}\ \emph {et~al.}(2016)\citenamefont
  {Tuyenbayev} \emph {et~al.}}]{CALTimeDependence}%
  \BibitemOpen
  \bibfield  {author} {\bibinfo {author} {\bibfnamefont {D.}~\bibnamefont
  {Tuyenbayev}} \emph {et~al.},\ }\href
  {http://iopscience.iop.org/article/10.1088/0264-9381/34/1/015002} {\bibfield
  {journal} {\bibinfo  {journal} {Classical Quant. Grav.}\ }\textbf {\bibinfo
  {volume} {34}},\ \bibinfo {pages} {015002} (\bibinfo {year}
  {2016})}\BibitemShut {NoStop}%
\bibitem [{\citenamefont {Hunter}(2007)}]{matplotlib}%
  \BibitemOpen
  \bibfield  {author} {\bibinfo {author} {\bibfnamefont {J.~D.}\ \bibnamefont
  {Hunter}},\ }\href {http://aip.scitation.org/doi/pdf/10.1109/MCSE.2007.55}
  {\bibfield  {journal} {\bibinfo  {journal} {Comput. Sci. Eng.}\ }\textbf
  {\bibinfo {volume} {9}},\ \bibinfo {pages} {90} (\bibinfo {year}
  {2007})}\BibitemShut {NoStop}%
\bibitem [{\citenamefont {Foreman-Mackey}\ \emph {et~al.}(2013)\citenamefont
  {Foreman-Mackey} \emph {et~al.}}]{emcee}%
  \BibitemOpen
  \bibfield  {author} {\bibinfo {author} {\bibfnamefont {D.}~\bibnamefont
  {Foreman-Mackey}} \emph {et~al.},\ }\href {https://arxiv.org/pdf/1202.3665}
  {\bibfield  {journal} {\bibinfo  {journal} {PASP}\ }\textbf {\bibinfo
  {volume} {125}},\ \bibinfo {pages} {306} (\bibinfo {year}
  {2013})}\BibitemShut {NoStop}%
\bibitem [{\citenamefont {Foreman-Mackey}(2016)}]{22}%
  \BibitemOpen
  \bibfield  {author} {\bibinfo {author} {\bibfnamefont {D.}~\bibnamefont
  {Foreman-Mackey}},\ }\href {\doibase 10.21105/joss.00024} {\bibfield
  {journal} {\bibinfo  {journal} {JOSS}\ }\textbf {\bibinfo {volume} {1}}
  (\bibinfo {year} {2016}),\ 10.21105/joss.00024}\BibitemShut {NoStop}%
\bibitem [{\citenamefont {Rasmussen}\ and\ \citenamefont
  {Williams}(2006)}]{Rasmussen2006}%
  \BibitemOpen
  \bibfield  {author} {\bibinfo {author} {\bibfnamefont {C.~E.}\ \bibnamefont
  {Rasmussen}}\ and\ \bibinfo {author} {\bibfnamefont {C.~K.~I.}\ \bibnamefont
  {Williams}},\ }\href
  {http://www.newton.ac.uk/files/seminar/20070809140015001-150844.pdf} {\emph
  {\bibinfo {title} {Gaussian processes for machine learning}}},\ Vol.~\bibinfo
  {volume} {1}\ (\bibinfo  {publisher} {MIT Press},\ \bibinfo {year}
  {2006})\BibitemShut {NoStop}%
\bibitem [{\citenamefont {Pedregosa}\ \emph {et~al.}(2011)\citenamefont
  {Pedregosa} \emph {et~al.}}]{sklearn}%
  \BibitemOpen
  \bibfield  {author} {\bibinfo {author} {\bibfnamefont {F.}~\bibnamefont
  {Pedregosa}} \emph {et~al.},\ }\href
  {http://www.jmlr.org/papers/v12/pedregosa11a.html} {\bibfield  {journal}
  {\bibinfo  {journal} {JMLR}\ }\textbf {\bibinfo {volume} {12}},\ \bibinfo
  {pages} {2825} (\bibinfo {year} {2011})}\BibitemShut {NoStop}%
\end{thebibliography}%

\end{document}